\documentclass[twocolumn,preprintnumbers,superscriptaddress,endnote,nofootinbib,aps,prd,floatfix]{revtex4}

\usepackage[utf8]{inputenc}

\usepackage{footmisc,enumerate}
\usepackage{multirow}
\usepackage{subfigure}
\usepackage{amsmath}
\usepackage{graphicx}
\usepackage{placeins}
\usepackage{xspace}
\usepackage{hyperref}
\hypersetup{colorlinks=true, citecolor=blue, urlcolor=blue, linkcolor=blue}
\usepackage[normalem]{ulem}

\renewcommand{\vec}[1]{{\bf{#1}}}

\newcommand{\Ref}[1]{Ref.~\cite{#1}}
\newcommand{\Eq}[1]{Eq.~\eqref{#1}}
\newcommand{\Fig}[1]{Fig.~\ref{#1}}

\newcommand{\Sec}[1]{Sec.~\ref{#1}}
\newcommand{\hjj}{\ensuremath{hjj}\xspace}
\newcommand{\tth}{\ensuremath{t\bar{t}h}\xspace}
\newcommand{\dphijj}{\ensuremath{\Delta\phi_{jj}}\xspace}
\newcommand{\dphill}{\ensuremath{\Delta\phi_{\ell\ell}}\xspace}
\newcommand{\dphiX}{\ensuremath{\Delta\phi_{X}}\xspace}
\newcommand{\cg}{\ensuremath{\tilde{c}_g}\xspace}
\newcommand{\ct}{\ensuremath{\tilde{c}_t}\xspace}
\newcommand{\tx}[1]{\text{#1}}
\newcommand{\dphi}[1]{\frac{\tx{d}\sigma_{#1}}{\tx{d}(\dphiX)}}
\newcommand{\gaga}{\gamma\gamma}
\newcommand{\Zga}{Z\gamma}
\newcommand{\GaSM}{\Gamma_{\text{SM}}}
\newcommand{\GaEFT}{\Gamma_{\text{dim.6}}}
\newcommand{\dthflat}{\delta_{\text{th}}}
\newcommand{\dsysflat}{\delta_{\text{sys}}^{\text{flat}}}
\newcommand{\dsysshape}{\delta_{\text{sys}}^{\text{shape}}}
\newcommand{\dsysflatproc}[1]{\delta_{\text{sys,}#1}^{\text{flat}}}
\newcommand{\dsysshapeproc}[1]{\delta_{\text{sys,}#1}^{\text{shape}}}
\newcommand{\dthproc}[1]{\delta_{\text{th,}#1}}

\begin{document}

\newcommand{\og}{\ensuremath{\tilde{O}_g}\xspace}
\newcommand{\ot}{\ensuremath{\tilde{O}_t}\xspace}

\preprint{IPPP/18/109}

\title{Approaching robust EFT limits for CP-violation in the Higgs sector}

\begin{abstract}
Constraining CP-violating interactions in effective field theory (EFT) of dimension six faces two challenges. Firstly, degeneracies in the multi-dimensional space of Wilson coefficients have to be lifted. Secondly, quadratic contributions of CP-odd dimension six operators are difficult to disentangle from squared contributions of CP-even dimension six operators and from linear contributions of dimension eight operators. Both of these problems are present when new sources of CP-violation are present in the interactions between the Higgs boson and heavy strongly-interacting fermions. We show that degeneracies in the Wilson coefficients can be removed by combining measurements of Higgs-plus-two-jet production via gluon fusion with measurements of top-pair associated Higgs production. In addition, we demonstrate that the sensitivity of the analysis can be improved by exploiting the top-quark threshold in the gluon fusion process. Finally, we substantiate a perturbative argument about the validity of EFT by comparing the quadratic and linear contributions from CP-odd dimension six operators and use this to show explicitly that high statistics measurements at future colliders enable the extraction of perturbatively robust constraints on the associated Wilson coefficients.
\end{abstract}

\author{Christoph Englert} \email{christoph.englert@glasgow.ac.uk}
\affiliation{SUPA, School of Physics \& Astronomy, University of Glasgow, Glasgow G12 8QQ, UK\\[0.1cm]}
\author{Peter Galler} \email{peter.galler@glasgow.ac.uk}
\affiliation{SUPA, School of Physics \& Astronomy, University of Glasgow, Glasgow G12 8QQ, UK\\[0.1cm]}
\author{Andrew Pilkington} \email{andrew.pilkington@manchester.ac.uk}
\affiliation{School of Physics \& Astronomy, University of Manchester, Manchester M13 9PL, UK\\[0.1cm]}
\author{Michael Spannowsky} \email{michael.spannowsky@durham.ac.uk}
\affiliation{Institute of Particle Physics Phenomenology, University of Durham, Durham DH1 3LE, UK\\[0.1cm]}

\pacs{}

\maketitle

\section{Introduction}
\label{sec:intro}
The search for new physics beyond the Standard Model (SM) is a central task of the Large Hadron Collider (LHC).
With established and motivated models under increasing pressure as more data get scrutinised, phenomenological
analyses have turned to largely model-independent measurement and interpretation strategies adopting the framework of SM
effective field theory (EFT)~\cite{Weinberg:1978kz,Buchmuller:1985jz,Burges:1983zg,Leung:1984ni,Hagiwara:1986vm}. SMEFT as a theoretical framework has undergone a rapid development over the past years, e.g.~\hbox{\cite{Jenkins:2013wua,Alonso:2013hga,Jenkins:2013zja,Elias-Miro:2014eia,Brivio:2017btx,deBlas:2017xtg,Buchalla:2017jlu,Helset:2018fgq}}.

EFTs facilitate the communication between the weak or measurement scale, and a UV completion that the EFT approach would
like to see itself contrasted with. As the UV completion of the SM is currently unknown, the leading operator dimension six deformations of the SM imply 2499 independent parameters~\cite{Grzadkowski:2010es} that should be considered as a priori free when we would like to constrain generic beyond the SM (BSM) physics that is sufficiently close to the decoupling limit to justify the dimension six approach.

Established phenomena such as the observed matter--anti-matter asymmetry, however, provide us a hint where motivated physics might 
be found, without making too many assumptions about the precise form of the UV completion itself. For instance, Sakharov's criteria~\cite{Sakharov:1967dj} of baryogenesis motivate the direct search for CP-violating effects in addition of the CP-violating sources in the SM, which are insufficient to account for the observed matter--anti-matter asymmetry. As the only source of CP violation in the SM is associated with the fermion-Higgs interactions, the Higgs sector naturally assumes a central role in such a search, in particular because its precise form is a lot less well-constrained compared to the gauge sectors. 

CP-violating effects associated with ``genuine'' dimension six effects, i.e. contributions that arise from the interference of SM contribution with dimension-six operators are limited to genuine CP-odd observables and asymmetries thereof~\cite{Brehmer:2017lrt,Bernlochner:2018opw} (see also~\cite{Ferreira:2017ymn}). In the context of Higgs physics, one motivated observable is the so-called signed $\phi_{jj}$~\cite{Hankele:2006ma} (see also~\cite{Plehn:2001nj,Klamke:2007cu,Campanario:2010mi,Campanario:2013mga,Campanario:2014oua}) in gluon and weak boson fusion. The first measurements of the signed $\phi_{jj}$ were recently published by the ATLAS Collaboration in the $h\rightarrow\gamma\gamma$ \cite{Aaboud:2018xdt} and $h\rightarrow ZZ$ \cite{Aaboud:2017oem} decay channels. An analogous observable can be constructed for top quark-associated Higgs production as well, as discussed in detail recently in Ref.~\cite{Khatibi:2014bsa,Goncalves:2018agy}. Working in the dimension six linearised approximation, such observables are the only phenomenologically viable ones because the interference terms cancel identically for any CP-even observable, such as total cross sections, decay widths as well as momentum transfer-dependent observables such as transverse momenta and invariant masses. 

Issues arise, however, when multiple operators affect the same observable. In this case, large CP-violating effects in two or more operators can completely cancel, yielding a result that resembles the SM prediction. Higgs-plus-two-jet production via gluon fusion receives corrections from heavy strongly-interacting fermions, including the top quark and possible as-yet-undiscovered heavy fermions that lie far above the electroweak scale. In the effective field theory approach, we can express this as corrections from two operators
\begin{equation}
\begin{split}
\label{eq:ops}
\og&={\alpha_s\over 8\pi v} G^{a}_{\mu\nu} \tilde{G}^{a\; \mu\nu} h,  \quad {\rm and} \\
\ot&=i \bar t \gamma_5  t h \,,
\end{split}
\end{equation}
where $t$ denotes the top quark, $G^{a}_{\mu\nu}$ is the gluon field strength with dual $\tilde G^{a\;\mu\nu} = \epsilon^{\mu\nu\rho\delta}G^a_{\rho\delta}/2$, $h$ represents the physical Higgs boson with mass $m_h=125~\text{GeV}$ and $v=246~\text{GeV}$ is the Higgs' vacuum expectation value.\footnote{The normalisation of $\og$ corresponds to integrating out the top quark with CP-odd couplings with Yukawa coupling size $\sqrt{2}m_t/v$ in the limit $m_t\to \infty$~\cite{Djouadi:1993ji,Kauffman:1993nv,Kniehl:1995tn}.} In the following we will denote $\cg,\ct$ as the corresponding Wilson coefficients of Eq.~\eqref{eq:ops}. While \og and \ot in Eq.~\eqref{eq:ops} are nominally of mass dimension five and four, respectively, they are derived from corresponding dimension six operators by appropriately replacing the Higgs field with its vacuum expectation value $v$ and setting the energy scale $\Lambda=v$. Additional contributions from the chromo-electric dipole moment can be constrained in, e.g., top production~\cite{Bernreuther:2015yna} and are not considered here.

It is well known that the $m_t$-associated threshold effects allow us to differentiate between these parameters in their CP-even manifestation using momentum transfer-dependent observables~\cite{Banfi:2013yoa,Grojean:2013nya,Azatov:2013xha,Schlaffer:2014osa,Buschmann:2014twa,Buschmann:2014sia,Azatov:2016xik,Englert:2018cfo}. Together with the information from top quark-associated Higgs production, this is enough to sufficiently disentangle the gluon-Higgs interactions from the top-Higgs contributions~\cite{Butter:2016cvz,Englert:2017aqb,Ellis:2018gqa}. In the case of the CP-odd operators of \Eq{eq:ops}, momentum transfer-dependent differential distributions used for the CP-even operators are identically zero for the new physics contribution. This makes the extraction of the CP-violating effects in the fermion-Higgs interactions and their separation from competing modifications of the gauge sector-Higgs interactions much more complicated to order $\sim \ct,\cg$.

The purpose of this work is to provide a detailed analysis of this issue and point out possible improvements that are straightforward to implement in existing experimental analyses. This paves the way to obtaining a more detailed picture of the Higgs CP properties at the LHC in the future.

Furthermore, we look at this analysis from the perspective of perturbative validity of the EFT approach. This is done by comparing linearised results to results obtained from including squared dimension six effects. The latter have been discussed in the past in detail (see e.g.~\cite{Plehn:2001nj,Hankele:2006ma,Englert:2012ct, Dolan:2014upa, Buckley:2015ctj,Dolan:2016qvg}). Yet, it is important to highlight that in this case any CP-even observable also acts as a probe of the CP-odd interactions. Hence, by including quadratic contributions searches and interpretations of CP-violating effects in the context of EFT become highly dependent on (often implicit) EFT assumptions. Our aim is to find experimental and phenomenological setups in which the quadratic contributions are negligible such that constraints on CP-violating interactions can be extracted perturbatively robust and with minimal assumptions on CP-even contributions. In the context of these considerations we extrapolate our analysis to experiments at future colliders.

Indirect constraints on $\ct$ have been derived e.g. in Ref.~\cite{Brod:2013cka} from electric dipole moments (EDMs).  Under the assumption that the Higgs couples to electrons with SM strength the most stringent constraints are obtained from measurements of the electron's EDM \cite{Andreev:2018ayy}, $\ct\sim\mathcal{O}(10^{-3})$.  In this analysis we focus on limits that can be obtained in direct measurements at the LHC and future colliders while investigating the interplay between \og and \ot as well as the perturbative behaviour of such limits.

This work is organised as follows: In Sec.~\ref{sec:setup} we outline our numerical setup and provide an overview of the relevant observables. We also place our analysis into the context of existing LHC analyses in the Higgs final states that we consider. We present our results in Sec.~\ref{sec:res}. In particular, we will comment on the comparison of dimension-six linearised approach with CP-even effects from CP-odd interactions as alluded to above and extrapolate our results to obtain LHC and future hadron collider projections. We conclude in Sec.~\ref{sec:conc}.

\section{Setup: Processes and Observables}
\label{sec:setup}

\subsection{Processes}
To analyse the prospects of discriminating $\og$ from $\ot$ via the process $pp\to hjj$, we use a modified version of {\sc{Vbfnlo}}~\hbox{\cite{Baglio:2014uba,Arnold:2008rz}}. Including dimension six interactions, we can write the full squared amplitude
\begin{equation}
\label{eq:ampsplit}
|{\cal{M}}|^2 = |{\cal{M}}_{\text{SM}}|^2 + 2\, \text{Re}\left( {\cal{M}}_{\text{SM}} {\cal{M}}_{\text{d6}}^\ast \right) 
+   |{\cal{M}}_{\text{d6}}|^2\,.
\end{equation}
Our modifications are such that the SM-interference and squared dimension six amplitude parts can be extracted individually, while
keeping the full top mass dependence of $\ot$~\cite{Hankele:2006ma,Klamke:2007cu,Campanario:2010mi,Campanario:2013mga}. The $\og$ contributions were tested against $\ot$ by approaching the $m_t\to \infty$ limit numerically. This provides a strong cross check of both implementations and our modifications, which is non-trivial by the fact that for linearised dimension six effects the integrated cross section is numerically zero (it is a CP-even observable), and genuine CP-sensitive observables need to be employed for such cross checks. 

\begin{figure*}[!t]
\subfigure[~]{\includegraphics[width=0.495\textwidth]{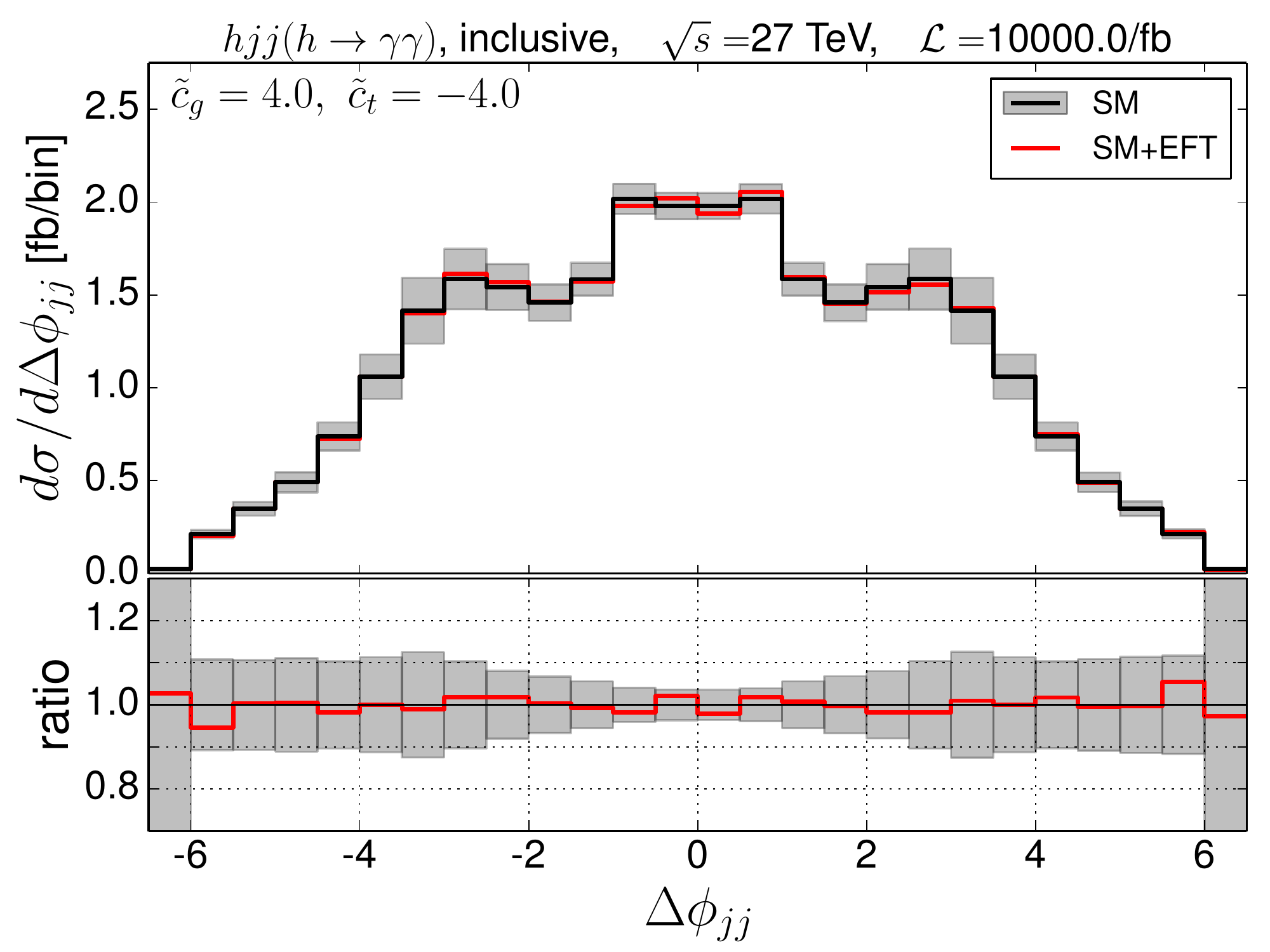}}
\subfigure[~]{\includegraphics[width=0.495\textwidth]{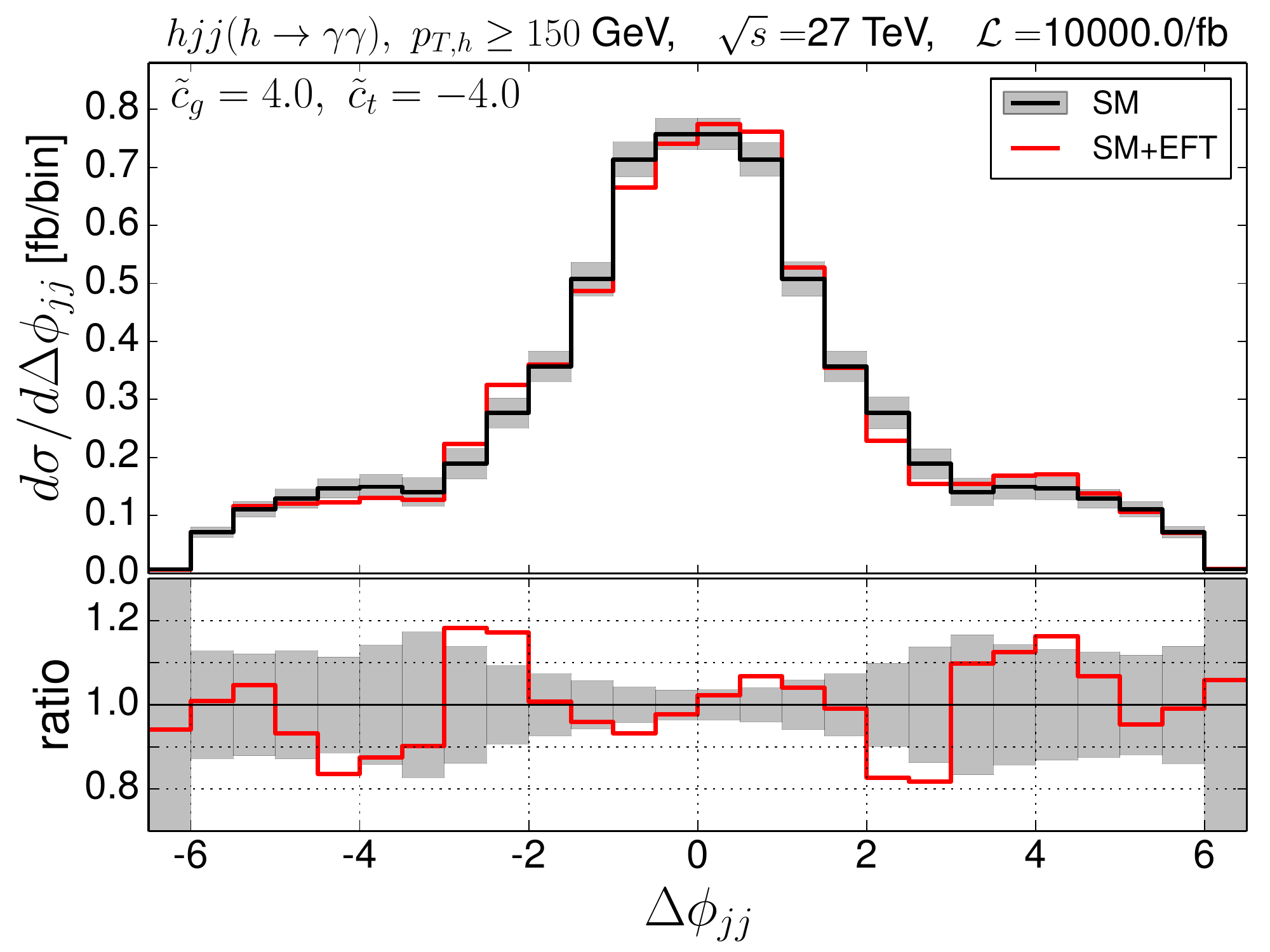}}
\caption{\label{fig:dphijjlin} $\dphijj$ distribution for \hjj production working in the linearised approximation ${\cal{O}}(\cg,\ct)$. Only the statistical uncertainty is shown in the plots.}
\end{figure*}

\begin{figure*}[!t]
\subfigure[~]{\includegraphics[width=0.495\textwidth]{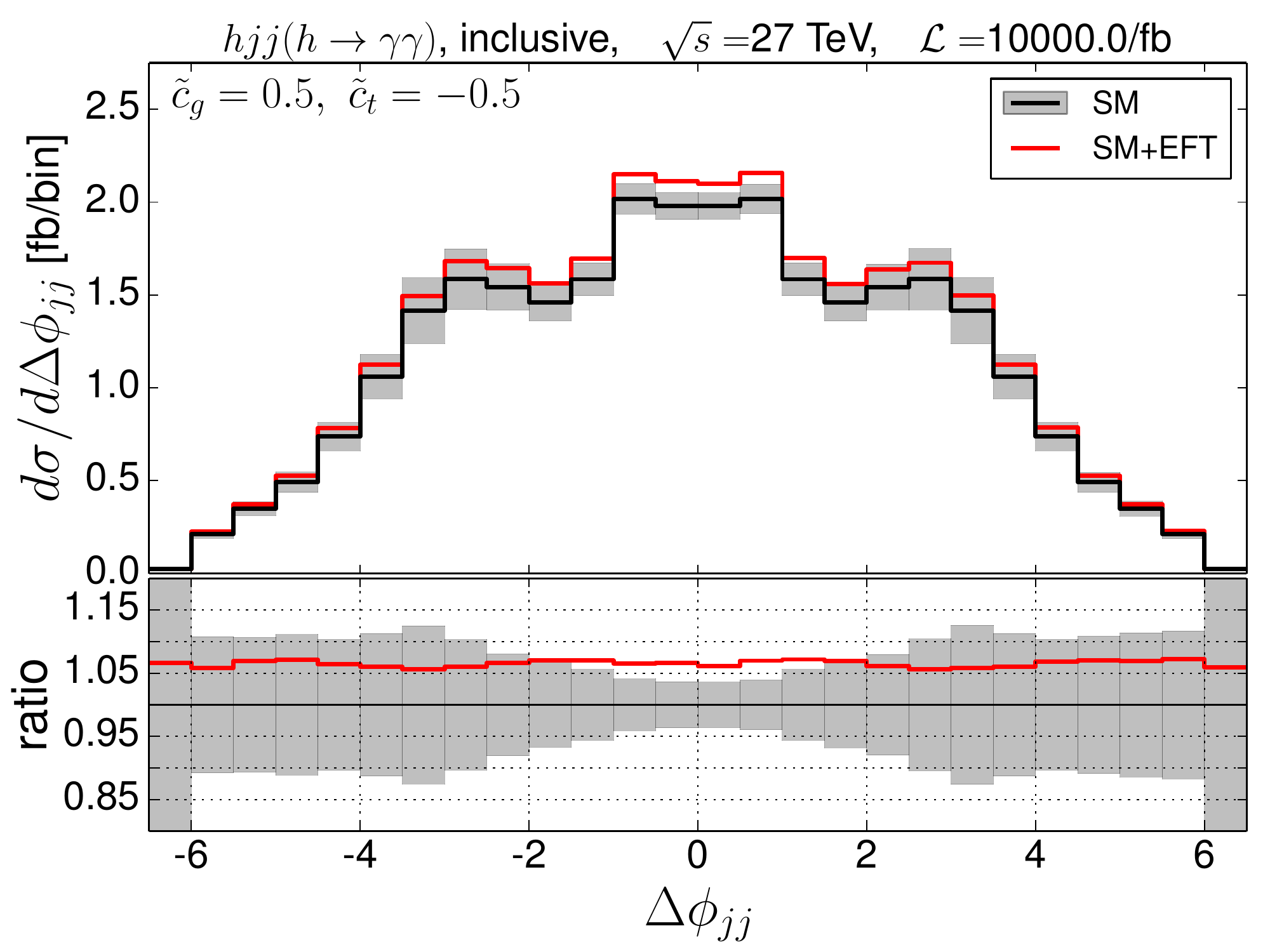}}
\subfigure[~]{\includegraphics[width=0.495\textwidth]{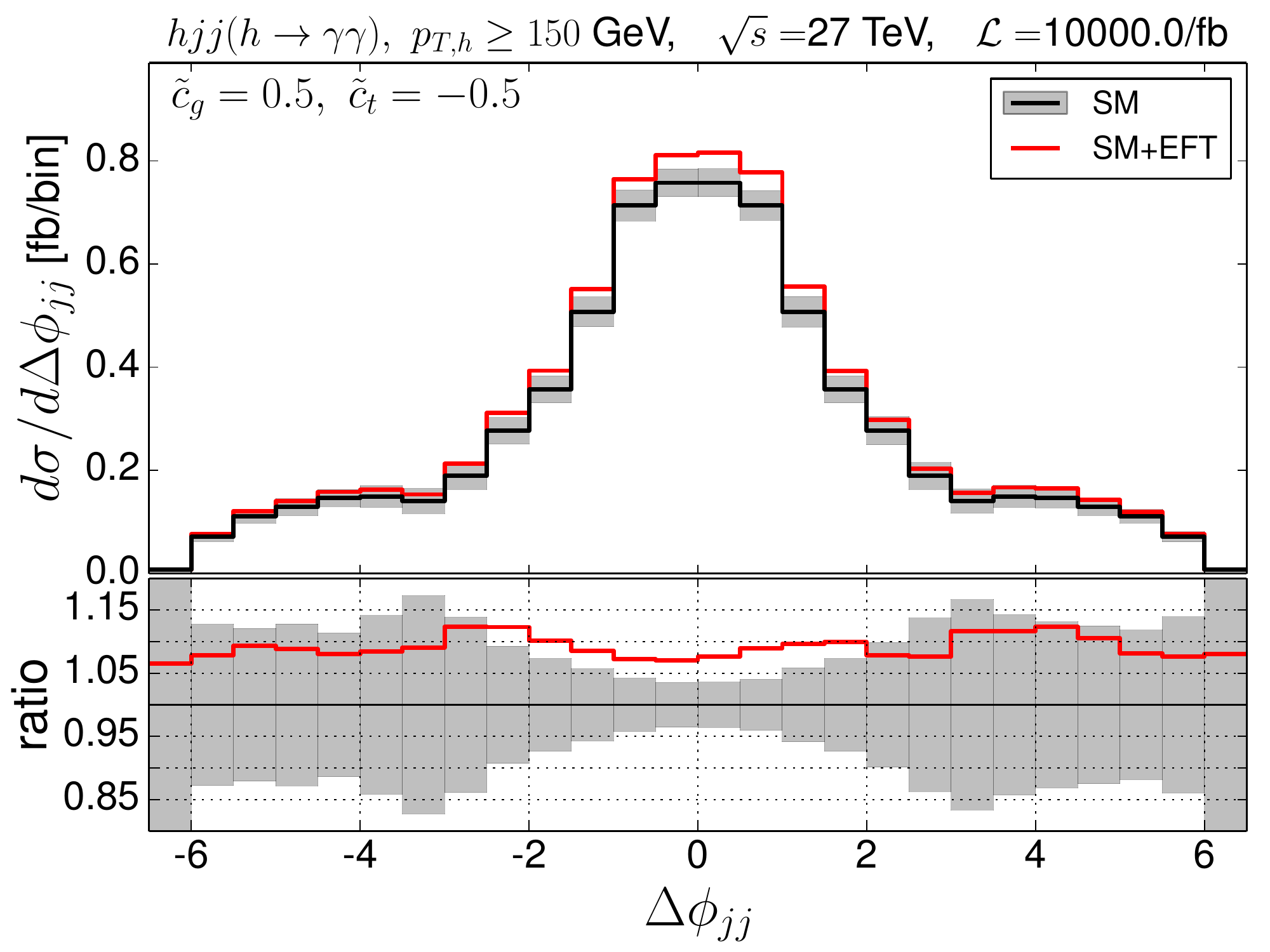}}
\caption{\label{fig:dphijjquad} $\dphijj$ distribution for \hjj production including quadratic contributions from the dimension six operators \og and \ot. Only the statistical uncertainty is shown in the plots.}
\end{figure*}

We output Les Houches events~\cite{Alwall:2006yp} of Higgs production in association with two light jets, \hjj and subsequently shower and hadronise them with {\sc{Herwig}}~\hbox{\cite{Bahr:2008pv,Bellm:2015jjp}}. For the analysis, we pass this output through a {\sc{Rivet}}~\cite{Buckley:2010ar} analysis which closely follows the event selection of~Ref.~\cite{Aaboud:2018xdt}. From the SM sample, we determine the event selection efficiencies on a bin-by-bin level by comparing parton-level with particle-level ({\sc{Rivet}}) analysis and Ref.~\cite{Aaboud:2018xdt}. We study the \hjj production channel in the $h\to\gaga$ decay mode with the possibility of including modifications of the Higgs branching ratios for comparisons (see below). We use flat K factors of 1.5 (13/27 TeV) and 1.18 (100 TeV) following Ref.~\cite{Contino:2016spe}. 

\begin{figure*}[!t]
\subfigure[~]{\includegraphics[width=0.495\textwidth]{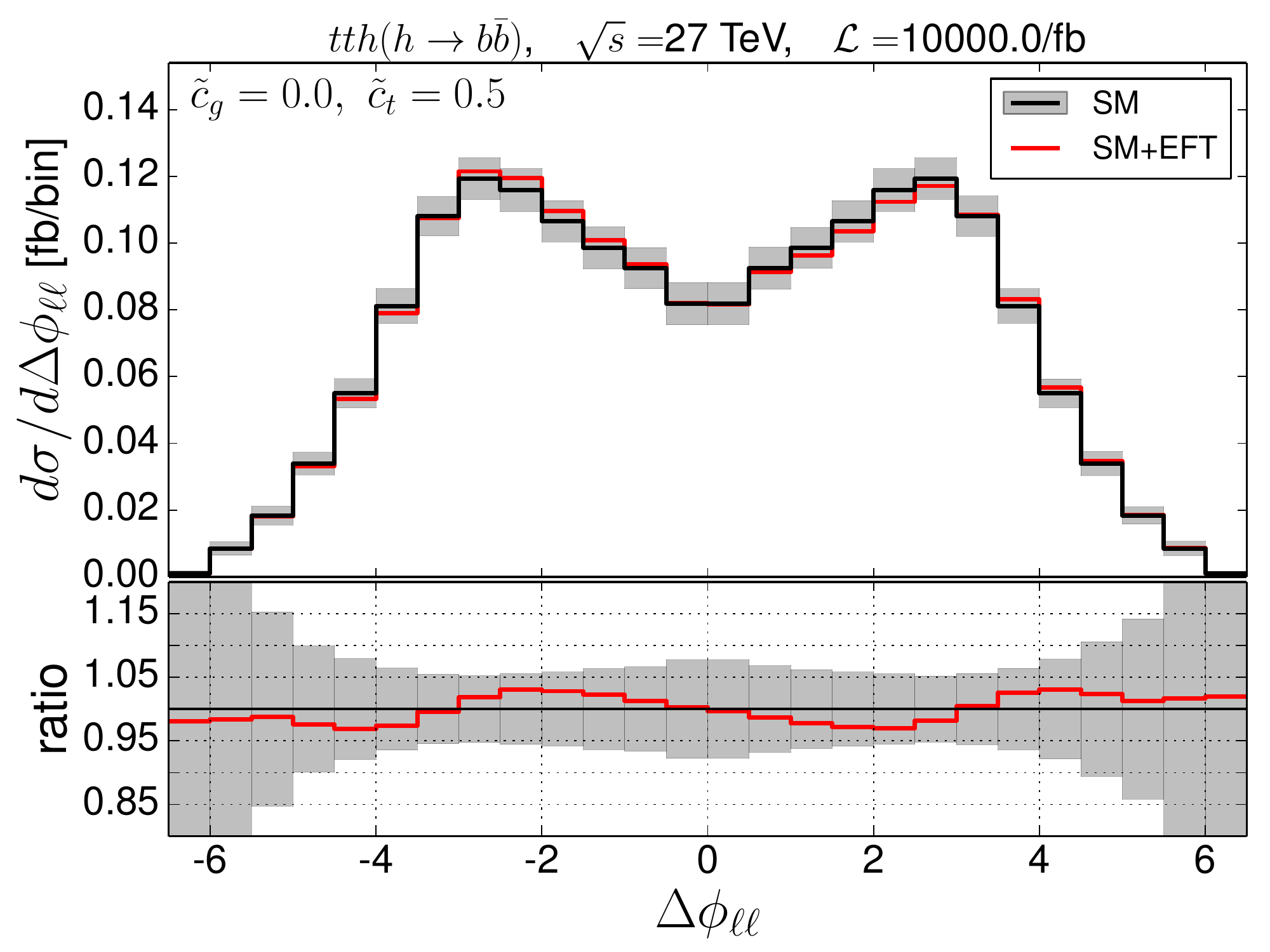}}
\subfigure[~]{\includegraphics[width=0.495\textwidth]{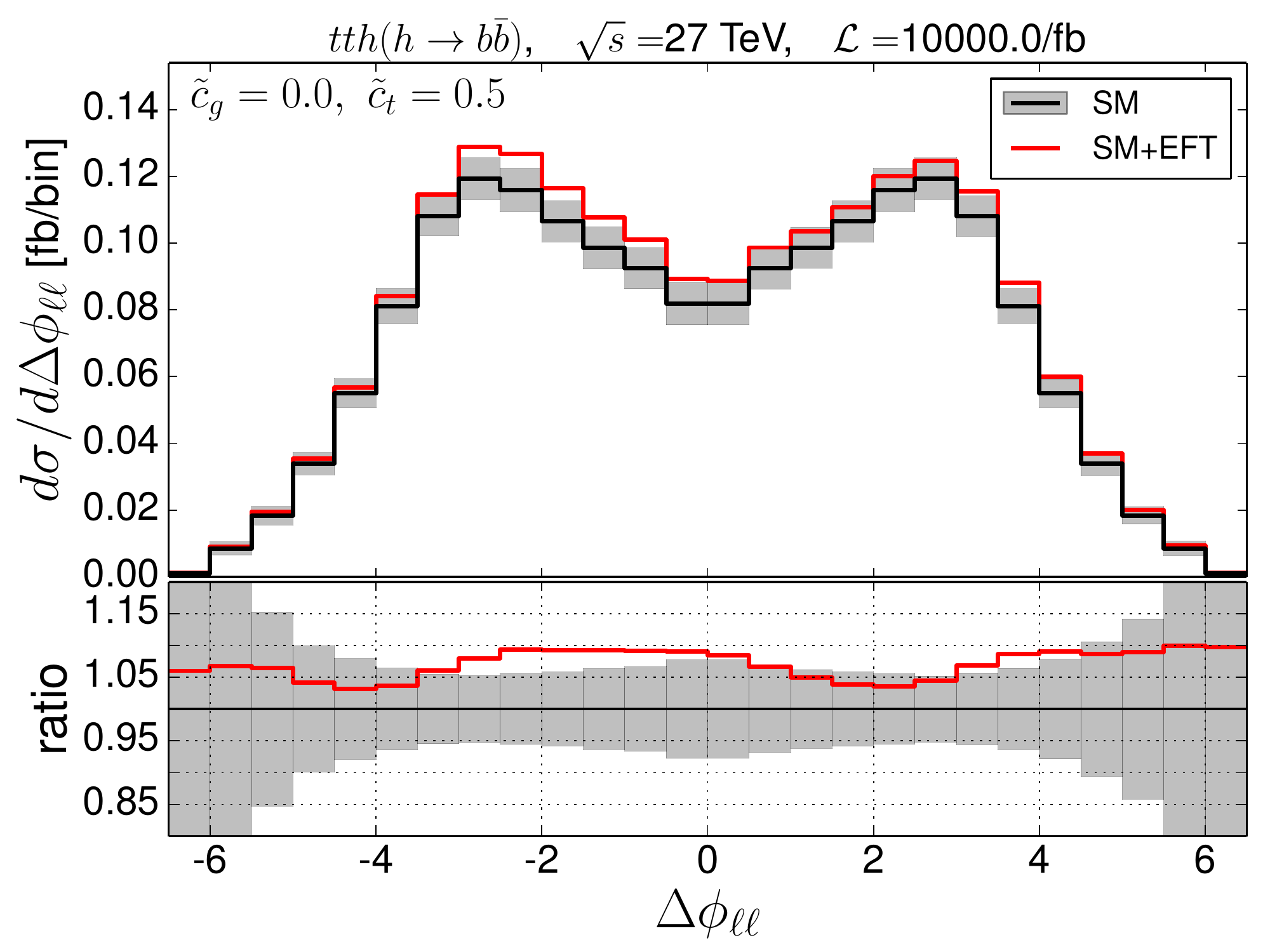}}
\caption{\label{fig:tthasym} $\dphill$ distribution for \tth production comparing linear (a) and quadratic (b) dimension six interactions. Only the statistical uncertainty is shown in the plots.}
\end{figure*}

To disentangle Higgs-gluon from Higgs-top interactions we also consider \tth events which are generated with \mbox{{\sc{MadGraph 5}}}~\cite{Alwall:2014hca} where the contribution from the effective operators in \Eq{eq:ops} have been implemented through a {\sc{UFO}}~\cite{Degrande:2011ua} model file which we generated using \mbox{{\sc{FeynRules}}}~\cite{Alloul:2013bka}. We study the $t\bar{t}$-associated Higgs production in the $h\to b\bar{b}$ decay mode whose branching ratio is indirectly affected by \og and \ot as well. We detail the computation of branching ratios in the appendix.
Similar to the \hjj case, we separate the \tth contributions according to Eq.~\eqref{eq:ampsplit}. We focus exclusively on the production couplings and do not include CP-sensitive information that can be obtained from Higgs decays, through, e.g. angular observables. Such a measurement would constrain the properties of the Higgs coupling to a particular final state particle and not the properties related to its production. To reflect the impact of higher-order QCD corrections we include flat K factors of 1.30 (13/27 TeV) and 1.36 (100 TeV)~\cite{Reina:2001sf,Beenakker:2002nc,Dawson:2003zu}\footnote{See also \url{https://twiki.cern.ch/twiki/bin/view/LHCPhysics/HiggsEuropeanStrategy}.}.

\subsection{Observables}
\label{subsec:obs}
Since we are interested in studying the CP-odd couplings of the Higgs to gluons and the top quark we study CP-sensitive observables. Therefore, in the \hjj channel, we calculate the signed azimuthal angle between the jets, which is defined as
\begin{equation}
\label{eq:dphijj}
\dphijj = \phi_{j,1} -\phi_{j,2}\,,
\end{equation}
where $\phi_{j,1}$ ($\phi_{j,2}$) is the azimuthal angle of the first (second) jet. The jets are ordered by their rapidity, i.e. 
\begin{equation}
y_{j,1}>y_{j,2}\,,
\end{equation}
which promotes this angular distribution to a P-sensitive observable.

In \Fig{fig:dphijjlin}, we show the $\dphijj$ differential distribution for the linear approximation using a particular choice of \cg and \ct as an example. \Fig{fig:dphijjlin}~(a) illustrates that the effects of \og and \ot can be very small even for large values of \cg and \ct in the vicinity of the blind direction $\ct\sim-\cg$ if inclusive observables are considered. As can be seen from \Fig{fig:dphijjlin}~(b), once $\dphijj$ is defined with an additional binning in a kinematic observable such as the transverse momentum of the Higgs\footnote{We consider the Higgs $p_T$ distribution in the following as means to resolve the top-threshold;  jet-$p_T$ distributions are less sensitive to $m_t$ threshold effects.} that focuses on more exclusive events around the top quark threshold and above ($p_{T,h}\ge150$~GeV), we can start disentangling the $\cg$ and $\ct$ directions. In principle, a fully-binned two-dimensional distribution $(\dphijj, p_{T,h})$ could be considered. However, this would come at the price of a large reduction in statistics and an enlarged statistical uncertainty. While we consider two search regions separated by a 150~GeV $p_{T,h}$ cut, the number of search regions, as well as their separation could be treated as tuning parameters in a more realistic analysis. The ratio  plot of \Fig{fig:dphijjlin}~(b) also shows that the linear EFT contribution to the distribution is asymmetric and that the integrated cross section vanishes.

\medskip

The qualitative behavior of \Fig{fig:dphijjlin} can be understood from the $p_{T,h}$ differential distributions of the CP-even operators. For momentum transfers that resolve the top-Higgs interactions (and $\ot$ accordingly) the effect relative to $\og$ should decrease as absorptive contributions of the top-loop are probed. The effects are not large, as can be expected from the  success of the $m_t\to \infty$ approximation for SM \hjj production~\cite{DelDuca:2001eu,DelDuca:2001fn,DelDuca:2003ba,DelDuca:2006hk,Campbell:2006xx}.

While \Fig{fig:dphijjlin} shows the $\dphijj$ distribution in the linearised approximation \Fig{fig:dphijjquad} presents an example for the case where the quadratic terms are included. As \Fig{fig:dphijjquad} shows employing a top threshold related kinematical cut improves on lifting the blind direction also in the case where quadratic contributions are included. In this example we have chosen smaller values for the Wilson coefficients than for the linearised case because the \cg--\ct resolving power for larger values is mostly driven by the total cross section which does not vanish for the quadratic EFT contribution. This behaviour can already be observed for the inclusive case in \Fig{fig:dphijjquad}~(a) where the ratio plot shows a slight offset in the EFT contribution with respect to the SM. In the high-$p_{T,h}$ sample in \Fig{fig:dphijjquad}~(b) the relative contribution of the linear dimension six part is increased with respect to the inclusive case.

\medskip
In complete analogy to \hjj, for the \tth channel we consider the dileptonic decay of the top-quark pair (assuming an event selection efficiency of 2.5\% \cite{Aaboud:2017rss}) and study the signed azimuthal angle \dphill between the two charged leptons defined as
\begin{equation}
\label{eq:dphill}
\dphill = \phi_{\ell,1} -\phi_{\ell,2}\,,
\end{equation}
where $\phi_{\ell,1}$ ($\phi_{\ell,2}$) is the azimuthal angle of the first (second) charged lepton~\cite{Goncalves:2018agy}. The leptons are ordered according to their rapidity, i.e. 
\begin{equation}
y_{\ell,1}>y_{\ell,2}\,.
\end{equation}
As leptons $\ell$ we consider electrons and muons from the decays $t\to bW \to b\ell^+\nu_{\ell}$ and $t\to bW\to b\tau^+\nu_{\tau}\to b\ell^+\nu_{\ell}\nu_{\tau}\bar{\nu}_{\tau}$ and the charge conjugated processes. The clean fully-leptonic final states of \tth production can possibly be augmented by semi-leptonic final states with appropriate jet-matching that removes the Higgs final states. We limit ourselves here to the clean final state as we can expect reconstruction to be feasible without relying on non-transparent multivariate techniques at the price of reduced statistics. Example $\dphill$ distributions for the linear approximation as well as including quadratic terms are shown in \Fig{fig:tthasym}. Note that, although, \tth receives corrections $\sim \cg$, these contributions solely arise from dressing the $gg\to t\bar t$ topologies with initial-state Higgs radiation. This renders \tth almost insensitive to $\og$.

\section{Analysis and Results}
\label{sec:res}
\subsection{EFT-Linearised Approximation}
\label{subsec:linear}
In the first part of the analysis we investigate the SM and interference contributions. The SM contributions to the considered Higgs production channels are CP-even while the interference contributions are CP-odd. Since the inclusive cross section is a CP-even observable the contribution from the interference part is exactly zero. The Higgs branching ratios are not affected along the same lines and we adopt the branching ratios of the Higgs Cross Section Working Group in the following~\cite{Heinemeyer:2013tqa}. 

To set limits on the CP-odd couplings in \Eq{eq:ops} we study the differential distribution
\begin{equation}
\frac{\tx{d}\sigma(\cg,\ct)}{\tx{d}\dphiX}=\frac{\tx{d}\sigma_{\tx{SM}}}{\tx{d}\dphiX}+\cg \frac{\tx{d}\sigma_g}{\tx{d}\dphiX}+\ct\frac{\tx{d}\sigma_t}{\tx{d}\dphiX}\,,
\end{equation}
where $X=jj,\ell\ell$ and $\sigma_g$ ($\sigma_t$) is specific to the operator \og (\ot) but is independent of the Wilson coefficient \cg (\ct) by construction. Since in this P-odd differential distribution the linear dependence $\sim \cg,\ct$ is non-vanishing, it is possible to scan the behavior $\tx{d}\sigma(\cg,\ct,\dphiX)/\tx{d} p_{T,h}$ to isolate the individual contributions $\sigma_g,\sigma_t$ through their characteristic momentum-dependencies.

To facilitate the limit setting in an adapted way, we can scan the entire parameter space by sampling only two points, $(\cg,\ct)=(1,0)$ and $(\cg,\ct)=(0,1)$, for each Higgs production channel. We then perform a fit on the basis of a $\chi^2$ of the differential distribution obtained from the three different data sets, \hjj with \mbox{$p_{T,h}<150$~GeV}, \hjj with \mbox{$p_{T,h}\ge150$~GeV} and \tth. The $\chi^2$ test statistic is given by
\begin{align}
&\chi^2(\cg,\ct)\nonumber\\
&=\left(b_{\tx{SM}}^i - b_{\tx{SM+D6}}^i(\cg,\ct)\right)V_{ij}^{-1}\left(b_{\tx{SM}}^j - b_{\tx{SM+D6}}^j(\cg,\ct)\right)\,,
\end{align}
where $b^i_{\tx{SM}}$ is the expected measurement in the $i$th bin assuming the SM is correct, and $b_{\tx{SM+D6}}^i(\cg,\ct)$ represents the theoretical prediction for specified values of the Wilson coefficients. $V_{ij}$ is the covariance matrix that accounts for theoretical and experimental uncertainties. For the statistical uncertainty in the $\Delta\phi_{jj}$ distribution in the $h\rightarrow\gamma\gamma$ channel, we take the uncertainty in the measured fiducial cross section for Higgs-plus-two-jet production at 13 TeV and 36/fb \cite{Aaboud:2018xdt} and redistribute this across the bins of the observable. This is then rescaled to the respective centre-of-mass energies and luminosities used in this analysis. For the statistical uncertainty in the $\Delta\phi_{\ell\ell}$ distribution in \tth production, we take the measured uncertainty of the dilepton channel in Ref.~\cite{Aaboud:2017rss}, redistribute this across the bins of the observable, and then rescale to the appropriate centre-of-mass energy and luminosity. We assume each systematic error to be fully correlated and adopt the following values
\begin{widetext}
\begin{center}
\begin{tabular}{l@{\hskip 0.1in}l@{\hskip 0.1in}l@{\hskip 0.1in}l}
$\Delta\phi_{\ell\ell}$: &$\dthflat=10$\%, & $\dsysflat=20$\% \cite{Aaboud:2017rss}, &$\dsysshape=1.0$\% \cite{ATLAS-CONF-2018-027},\\[0.2cm]
$\Delta\phi_{jj}$: &$\dthflat=10$\% \cite{Aaboud:2018xdt},  & $\dsysflat=10$\% \cite{Aaboud:2018xdt},& $\dsysshape=2.5$\% \cite{Aaboud:2018xdt},
\end{tabular}
\end{center} 
\end{widetext}
where $\dthflat$ is the theoretical uncertainty in the fiducial cross section of each process, and $\dsysflat$ and $\dsysshape$ represent experimental uncertainties in the normalisation and shape of the expected measurements, respectively. Note that for \tth\ production, the current theoretical uncertainty due to background mismodelling in the experimental analysis is much larger than 10\%. However, we assume that this will be reduced in future analyses, due to improved theoretical models and increasing use of control regions with the larger datasets.

\subsection{Including Quadratic Dimension Six Terms}
\label{subsec:quad}
In the second part of the analysis we include the quadratic contributions. Analogously to the linear case we use the \dphiX distributions to calculate the $\chi^2$. Including quadratic contributions the \dphiX distributions are given by
\begin{multline}
\frac{\tx{d}\sigma(\cg,\ct)}{\tx{d}(\dphiX)}  = \dphi{\tx{SM}}+\cg\dphi{g}+\ct\dphi{t}\\
 + \cg\ct\dphi{gt}+\cg^2\dphi{g2}+\ct^2\dphi{t2}\,,
\end{multline}
i.e. we have to sample five parameter points per channel in order to scan the entire parameter space. Choosing $(\cg,\ct)=(1,-1)$ for the $\sigma_{gt}$ sample provides results with larger numerical stability as the histogram is sampled close to
the blind direction in \cg-\ct space and therefore gives only a small contribution from the dimension six operators.

\subsection{EFT validity}
EFT deformations of the SM typically lead to momentum enhancements which in turn can violate unitarity. Similar to the discussion of unitarity in the SM, where bounds on the Higgs mass were set by investigating partial waves, one would not interpret a Higgs mass in excess of unitarity constraints as a signal of breakdown of quantum mechanics in nature but as the indication that we deal with a strongly-coupled scenario where perturbative techniques are not justified (the Hamiltonian is real implying a unitary time evolution). As we ultimately rely on perturbative techniques in the simulation chain, there would be the need to assign a large uncertainty to the leading order approximation that would be related to the scale at which we probe the theory. As with all scale choices these are largely ad-hoc. A similar problem arises in EFT deformations where we can expect tails of energy-dependent distributions to be kinematically enhanced $\sim \vec{p}^2/\Lambda^2$. Separating off the $\Lambda$ dependence, Eq.~\eqref{eq:ampsplit} can then be phrased as
\begin{equation}
\label{eq:mod}
|{\cal{M}}|^2 = |{\cal{A}}_{\text{SM}}|^2 {1\over \Lambda^0} + 2\, \text{Re}\left( {\cal{A}}_{\text{SM}} {\cal{A}}_{\text{d6}}^\ast \right) {c\, Q^2 \over \Lambda^2}
+   |{\cal{A}}_{\text{d6}}|^2 {c^2\, Q^4 \over \Lambda^4}\,.
\end{equation}
Eq.~\eqref{eq:mod} directly singles out the stringent region 
\begin{equation}
\label{eq:val1}
|c | < {Q^2 \over \Lambda^2}\,.
\end{equation}
As we are considering average energy scales we choose a less stringent criterion 
\begin{equation}
\label{eq:val}
|c|{\langle Q\rangle \over v} \gtrsim {c^2} {\langle \tilde Q(c)\rangle^2 \over v^2} \;,
\end{equation}
where $Q$ and $\tilde Q$ are the median scales probed in the process, i.e. $Q$ corresponds to the average energy probed by $\text{Re}({\cal{M}}_{\text{SM}} {\cal{M}}_{\text{d6}}^\ast)$ and $\tilde Q$ is probed by $|{\cal{M}}_{\text{d6}}|^2$. 
We will comment on this choice and how it relates to the size of possible deviations from the SM as well as on differences to Eq.~\eqref{eq:val1} below.

Note that since $Q,\tilde Q$ are CP-even quantities, there is no dependence of $Q$ on the Wilson coefficients. We can therefore write
\begin{equation}
\langle Q\rangle = \langle \tilde Q(0)\rangle\,.
\end{equation}
The respective matrix element distributions sample the probed energy scales without making reference to the statistical sampling of the energy scales. Note that similar to using renormalisation and factorisation scales as measures to quantify associated uncertainties, this choice is ad-hoc and more constraining criteria can be formulated. 

While such a scaling is a typical behavior of perturbative models, we can expect it to be violated for non-perturbative
SM extensions. For the latter models, the naive hierarchy between dimension six and higher dimensional operators will be violated once we move closer
to the characteristic energy scale of the strong interactions, which signals the need to transition from the effective picture to the new relevant microscopic degrees of freedom (see also \cite{Contino:2016jqw,Alte:2018nbn} related discussions). Put differently, Eq.~\eqref{eq:val} is
a reason why we typically do not see large CP-violating phases in perturbative scenarios like two-Higgs doublet
models or the (N)MSSM.

On a more practical level, as we need to employ Monte Carlo techniques to simulate LHC final states that make use of fixed-order perturbation theory, our phenomenological modelling of a particular scenario cannot be trusted when Eq.~\eqref{eq:val} is badly violated. Rearranging leads straightforwardly to
\begin{equation}
|c| \lesssim {v \over \langle \tilde Q(c) \rangle} {\langle Q\rangle \over \langle \tilde Q(c)\rangle }\,.
\end{equation}
Numerically, we find approximate linear dependencies of the average probed scales as
\begin{alignat}{5}
\hbox{13~TeV}:\quad && \langle \tilde Q(c) \rangle \simeq 1077~\text{GeV} + 31.45 ~\text{GeV}\,|\cg| \nonumber\\
\hbox{27~TeV}:\quad && \langle \tilde Q(c) \rangle \simeq 1463~\text{GeV} + 89.48 ~\text{GeV}\,|\cg| \\
\hbox{100~TeV}:\quad && \langle \tilde Q(c) \rangle \simeq 2374~\text{GeV} + 386.57 ~\text{GeV}\,|\cg| \nonumber
\end{alignat}
i.e. the average probed energy does not depend too much on the size of the Wilson coefficient. For $\ct$ the dependence is
flat around the SM expectation. This is expected, as the modifications of $y_t$ only shift the cross section uniformly compared to the SM, which does not affect the sampling of the associated average probed energy. 

We find values
\begin{equation}
\label{eq:valid}
\begin{split}
13~\text{TeV}:\quad &  (|\cg|,|\ct|) \lesssim (0.23,0.23)\,,\\
27~\text{TeV}:\quad &  (|\cg|,|\ct|) \lesssim (0.16,0.17)\,,\\
100~\text{TeV}:\quad & (|\cg|,|\ct|) \lesssim (0.10,0.10)\,,\\
\end{split}
\end{equation}
i.e. under the criteria of Eq.~\eqref{eq:val} we can expect BSM contributions in the vicinity of $\lesssim 20\%$ compared to the SM. Again this is a typical ballpark of perturbative SM UV-completions. 
The more stringent requirement of
Eq.~\eqref{eq:val1} singles out regions of 5\%, 3\% and 1\% deviations from the SM at the 13, 27, and 100~TeV, respectively. Such deviations are hard to observe in the light of expected uncertainties and are pessimistic: If this criterion is adopted, our analysis does exhibit sensitivity in this region.

We find to good approximation
\begin{equation}
|c| \lesssim {v \over \langle Q\rangle} 
\end{equation}
it becomes clear that the range of the Wilson coefficients are quickly pushed to small values if the probed energy scale that characterises consistency with the SM is pushed to high values. Also if the new scale of physics (chosen earlier as $\Lambda=v$ for bookkeeping) lies far above the average probed scale, i.e. $\Lambda\gg \langle Q\rangle$ we see that the Wilson coefficient will only be loosely constrained. In this case the convergence of Eq.~\eqref{eq:mod} is guaranteed by 
the size of $\Lambda$ compared to kinematically relevant scales of the process, 
\begin{equation}
{c  \,Q^2\over\Lambda^2} \gg {c^2 \, Q^4\over \Lambda^4}
\end{equation}
but it is increasingly unlikely that the constraint will be important in UV matching calculations which translate the observed constraint into (perturbative) constraints of the UV parameter space.

\subsection{Results and Comparison}

\begin{figure}[t!]
\begin{center}
\includegraphics[height=6.5cm]{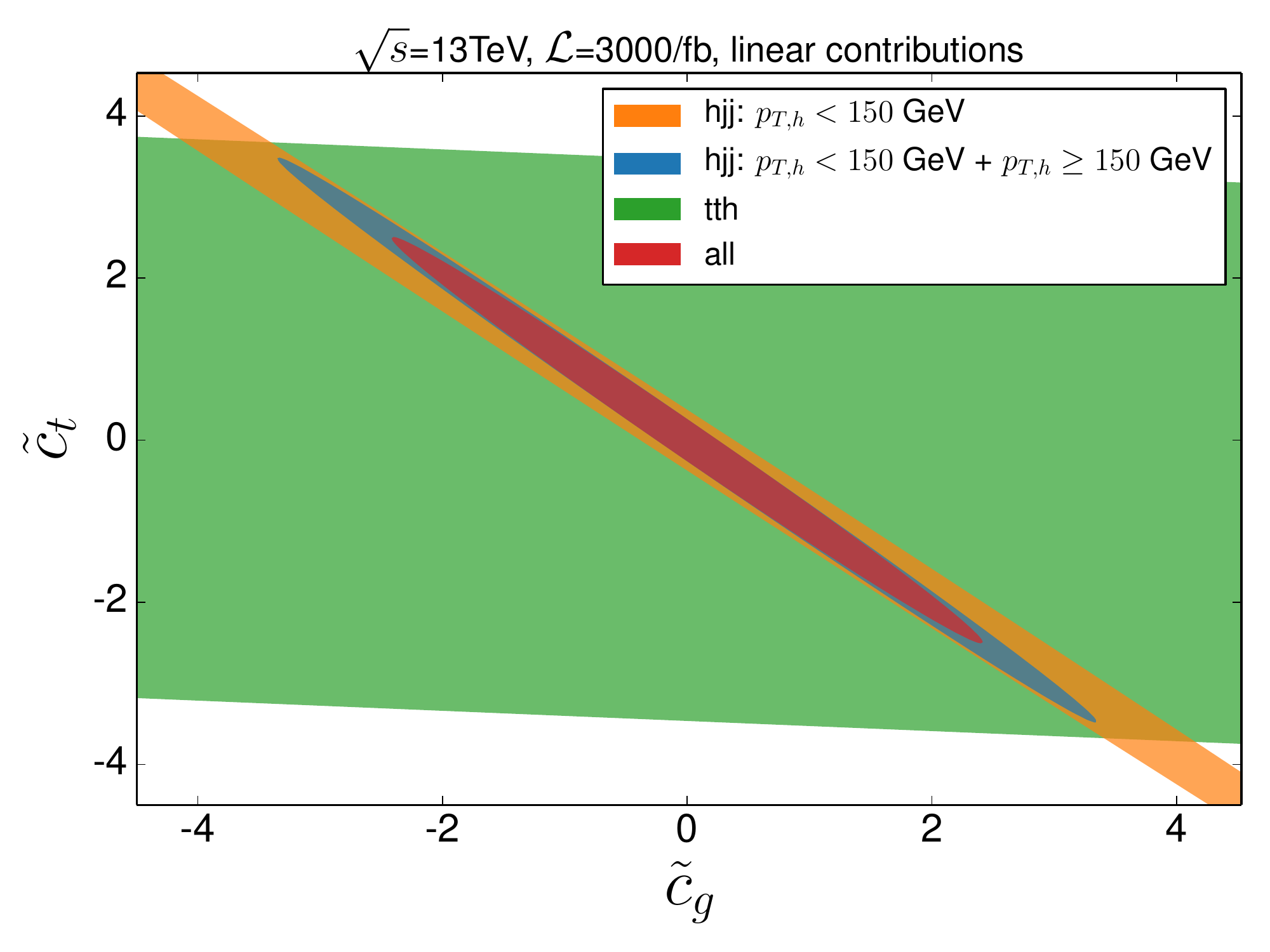}
\caption{Constraints for 95\% confidence level on \cg and \ct from different data sets at 13 TeV using the full \dphiX\ distributions, with $X=jj,\,\ell\ell$. Orange: only low-$p_{T,h}$ in \hjj, blue: combination of low $p_{T,h}$ and high $p_{T,h}$ events in \hjj, green: constraint from the \tth sample only, red: combination of all samples. In this plot only the contributions from SM and interference are taken into account.}
\label{fig:lin}
\end{center}
\end{figure}

\begin{figure}[t!]
\begin{center}
\includegraphics[height=6.5cm]{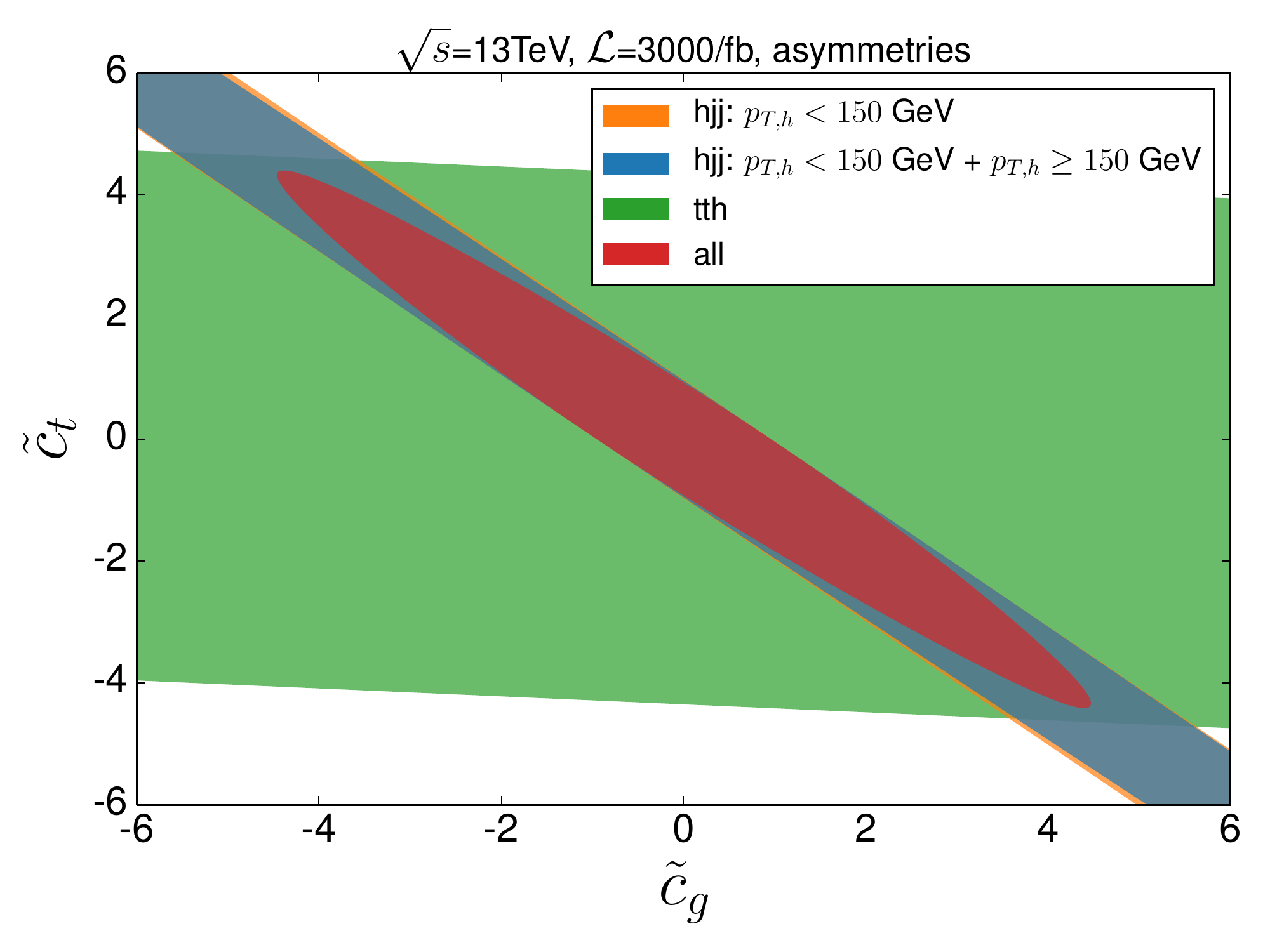}
\caption{Constraints for 95\% confidence level on \cg and \ct from different data sets at 13 TeV using simple asymmetries (Eq.~\ref{eq:asym}). Orange: only low-$p_{T,h}$ in \hjj, blue: combination of low $p_{T,h}$ and high $p_{T,h}$ events in \hjj, green: constraint from the \tth sample only, red: combination of all samples. In this plot only the contributions from SM and interference are taken into account.}
\label{fig:asym}
\end{center}
\end{figure}
\begin{figure}[t!]
\begin{center}
\includegraphics[height=6.5cm]{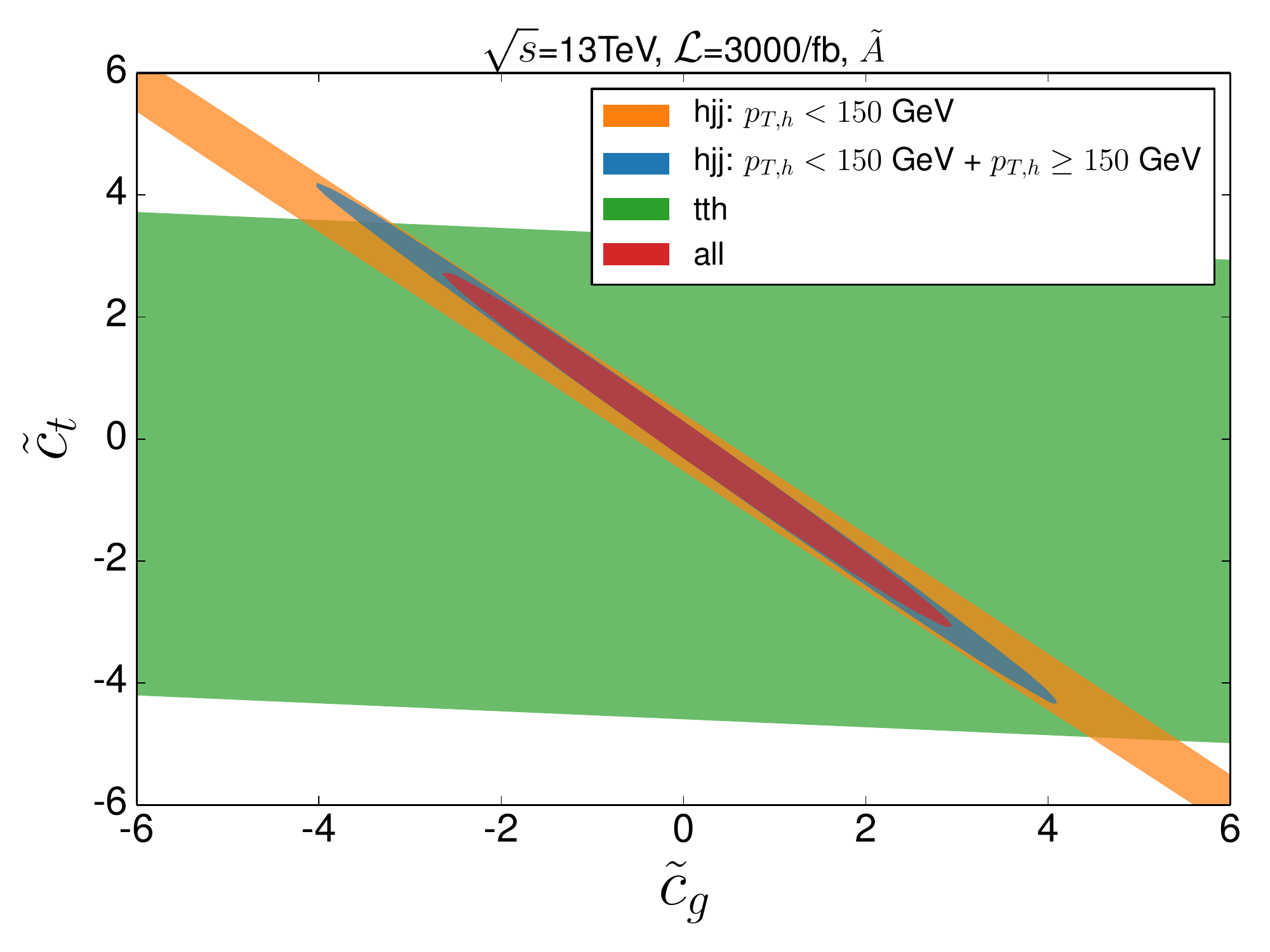}
\caption{Constraints for 95\% confidence level on \cg and \ct from different data sets at 13 TeV using binned asymmetries (Eq.~\ref{eq:absasym}). Orange: only low-$p_{T,h}$ in \hjj, blue: combination of low $p_{T,h}$ and high $p_{T,h}$ events in \hjj, green: constraint from the \tth sample only, red: combination of all samples. In this plot only the contributions from SM and interference are taken into account.}
\label{fig:absasym_13tev}
\end{center}
\end{figure}

\begin{figure}[t!]
\includegraphics[height=6.5cm]{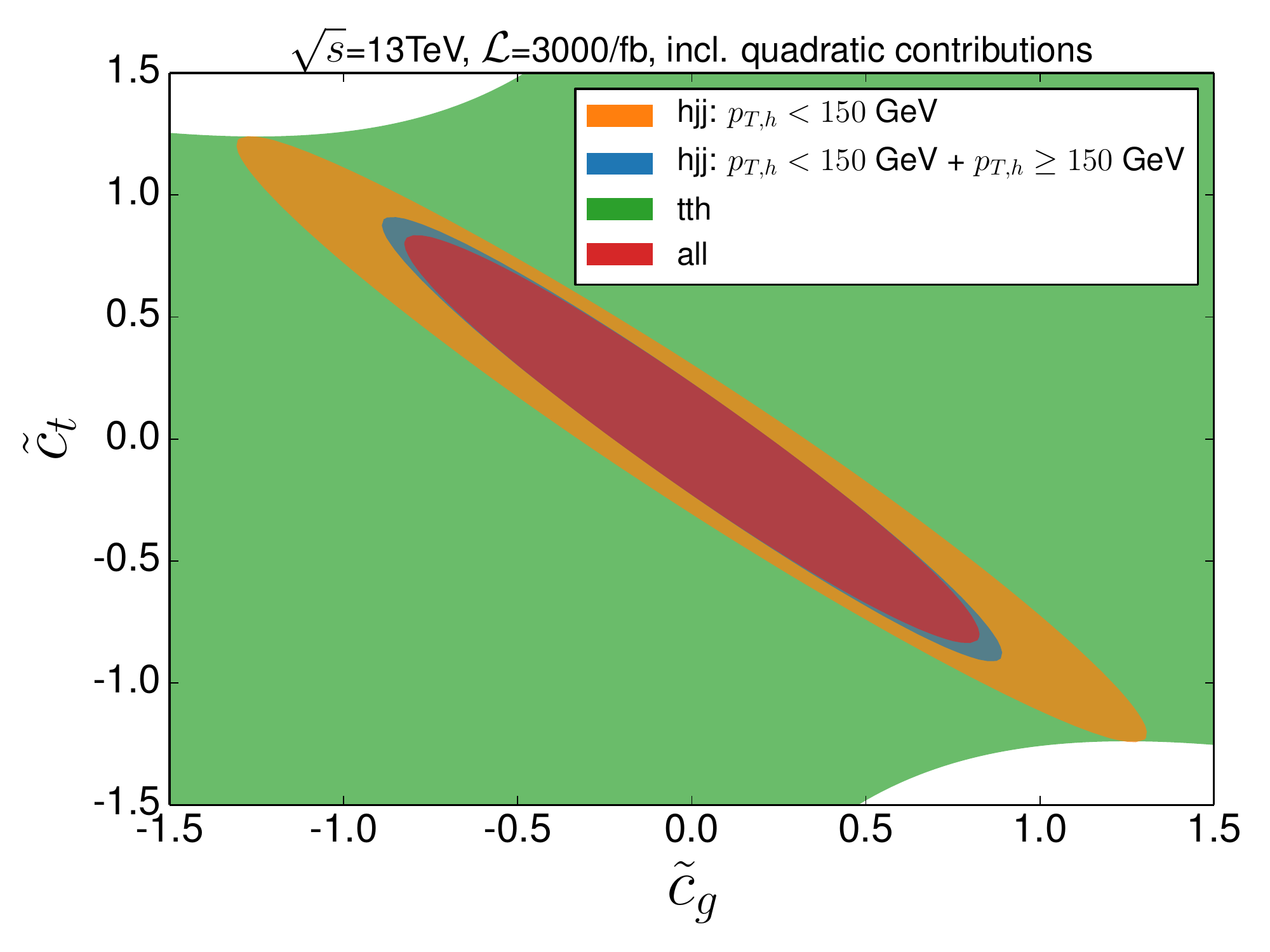}
\caption{Same as \Fig{fig:lin} but here the quadratic contributions are included as well.}
\label{fig:quad}
\end{figure}

\begin{figure}[t!]
\includegraphics[height=6.5cm]{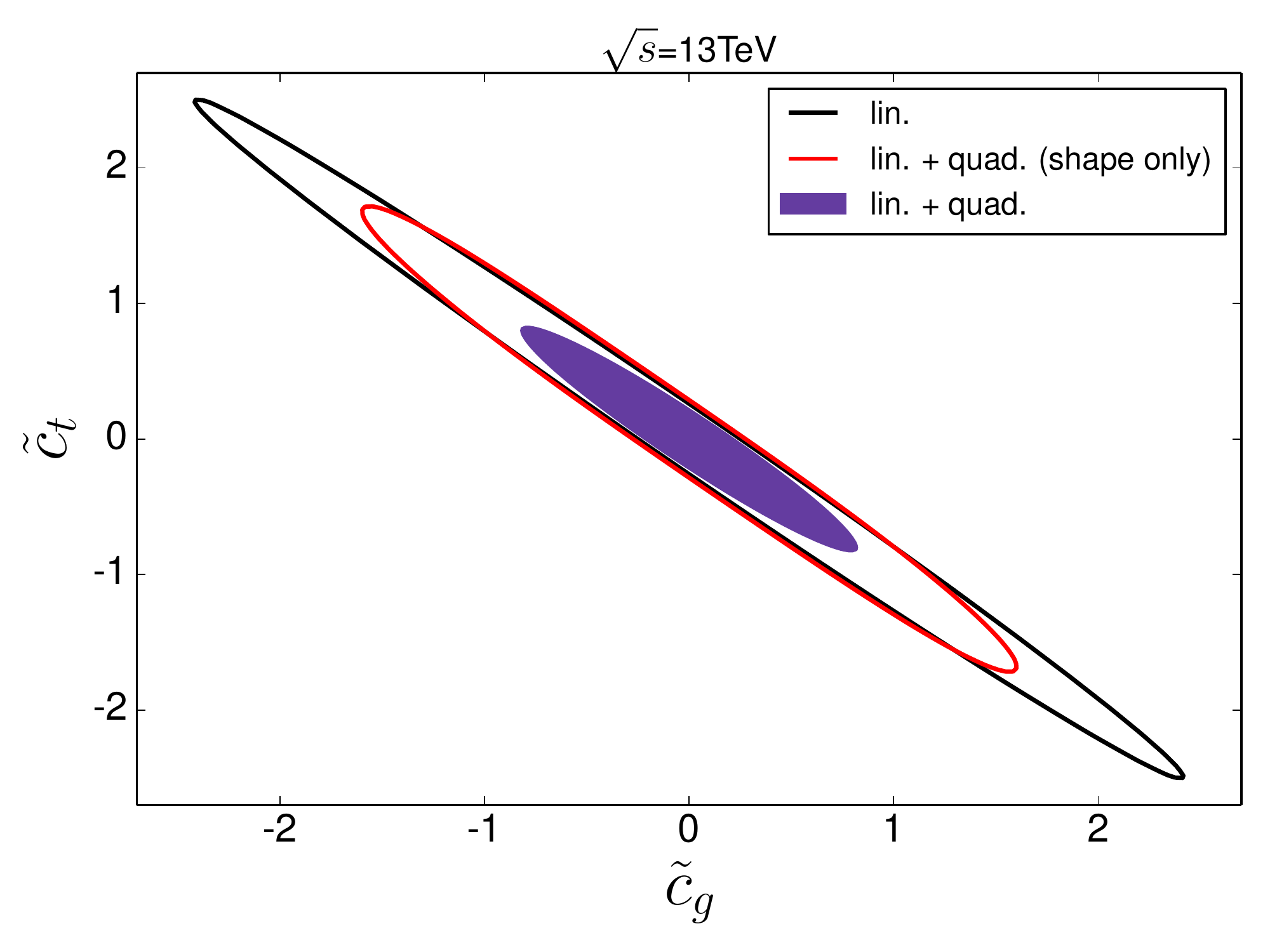}\hfill
\caption{\label{fig:linvsquadLHC}Comparison between the 95\% confidence level constraints on \cg and \ct obtain from linear dimension six contributions vs. linear and quadratic contributions at the LHC with $\sqrt{s}=13$ TeV and $\mathcal{L}=3000$/fb. The red ellipse shows the constraints for the linear and quadratic contributions when only the shape is used by normalizing the distributions to the SM cross section.}
\end{figure}

\begin{figure*}[t!]
\includegraphics[height=6.5cm]{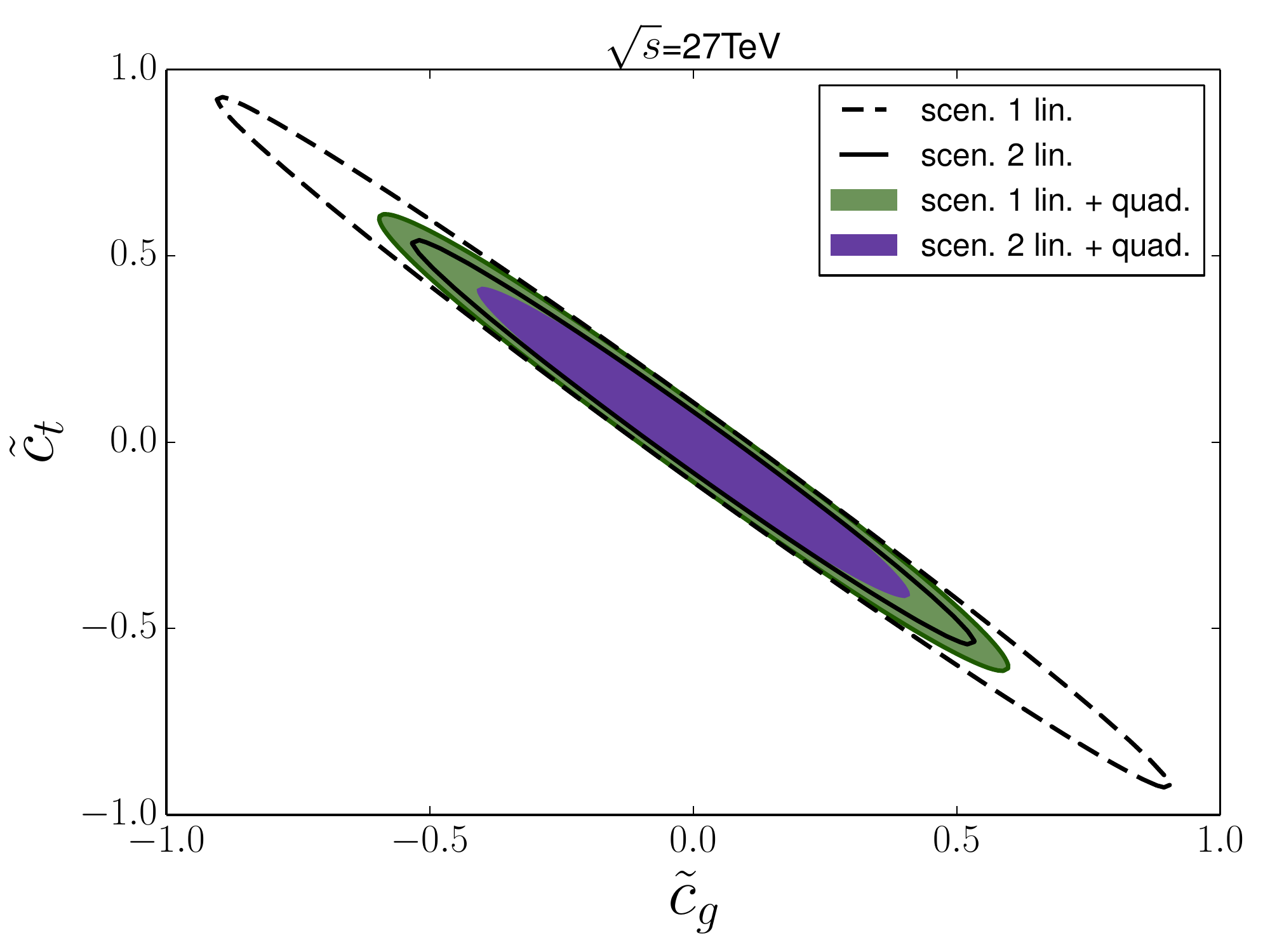}\hfill
\includegraphics[height=6.5cm]{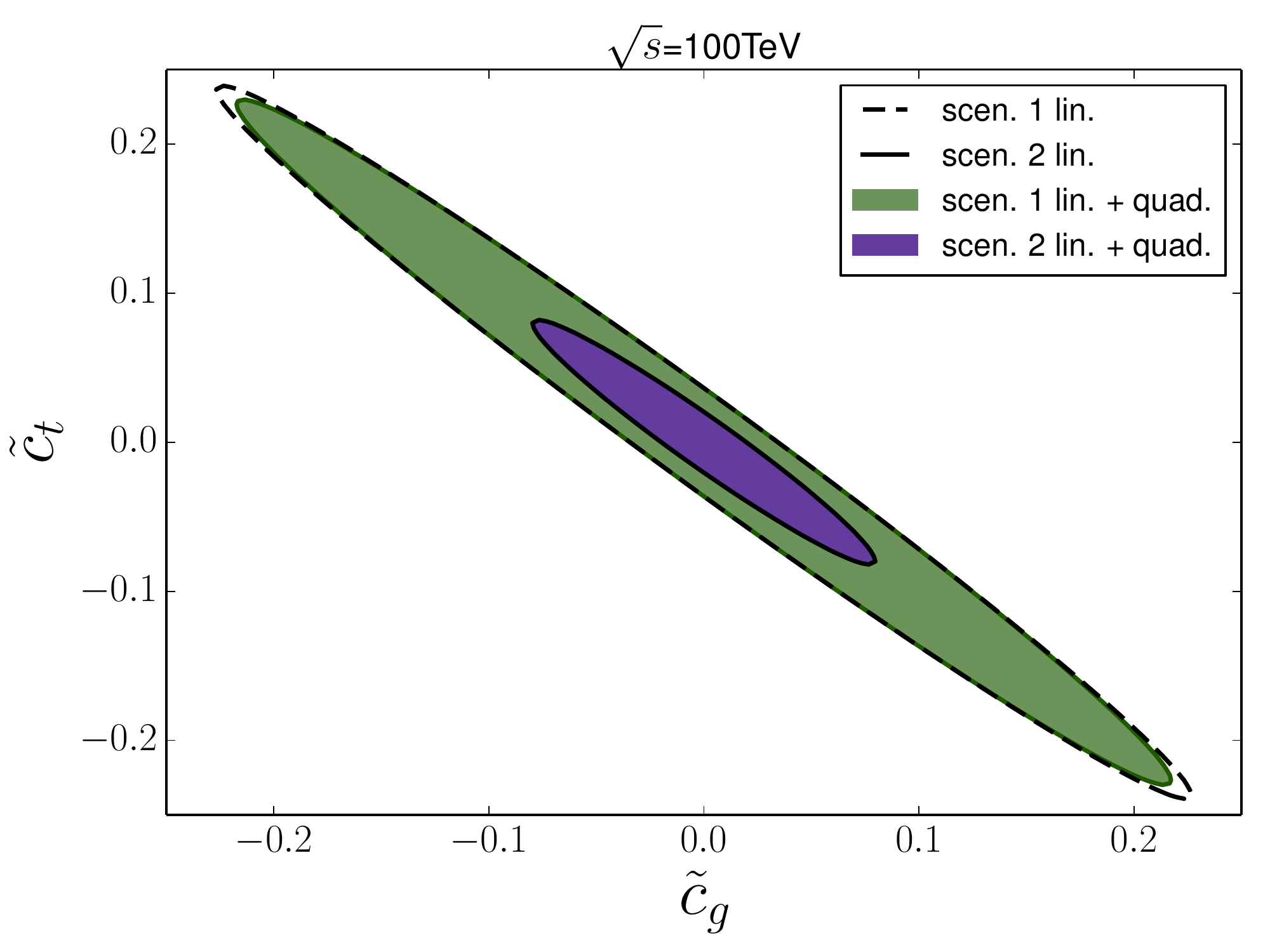}\hfill
\caption{\label{fig:comp_energies}Comparison between scenario 1 and 2 as well as linear and quadratic contributions of the 95\% confidence level constraints on \cg and \ct for 27 TeV and 100 TeV.}
\end{figure*}

In \Fig{fig:lin} the constraints on \cg and \ct at 95\% confidence level (CL) are presented for the LHC at 13 TeV showing the contributions from different production channels and kinematical regions. These contours are obtained by including only the SM and linear dimension six contributions. The green band in \Fig{fig:lin} represents the constraints using only the $t\bar{t}h$ sample. It illustrates that this channel is only very weakly sensitive to contributions from \cg as mentioned in \Sec{subsec:obs}. The slicing according to $p_T$, however, lifts the blind direction visible in the $hjj$ samples (orange and blue bands) to some extent. Comparing \Fig{fig:lin} with the perturbative bounds in \Eq{eq:valid} shows that generic searches for (C)P violation in production processes in the Higgs and top sector will be difficult at the LHC using only the decay modes considered. It will be necessary to use additional decay channels and to increase the dataset by combining results from both ATLAS and CMS.

One might raise the concern that CP-even contributions could contaminate these distributions such that it becomes difficult to disentangle the genuine CP-odd contributions of \og and \ot from those CP-even contributions. As an alternative strategy for the analysis of the linear dimension six contributions we also studied the asymmetries
\begin{equation}
A_X = \frac{\sigma(\dphiX<0)-\sigma(\dphiX>0)}{\sigma(\dphiX<0)+\sigma(\dphiX>0)}
\label{eq:asym}
\end{equation}
with $X=jj, \ell\ell$. For contributions symmetric in \dphiX, i.e. CP-even contributions, $A_X=0$. However, these asymmetries provide weaker constraints on \cg and \ct than the full distributions as can be seen in \Fig{fig:asym}. To recover the sensitivity of the full distributions, we propose the use of binned asymmetries defined as follows:
\begin{equation}
\tilde{A}_X = \frac{\sum_{i=1}^{N/2}\left|\sigma_i(\dphiX)-\sigma_{N-i+1}(\dphiX)\right|}{\sigma}\,,
\label{eq:absasym}
\end{equation}
where $X=jj,\,\ell\ell$, $\sigma_i(\dphiX)$ is the $i$th bin in the $\dphiX$ distribution and $N$ the number of bins. This definition assumes that $\dphiX<0$ for all bins $1\le i\le N/2$ and $\dphiX > 0$ for all bins $N/2+1\le i\le N$ and that the binning is symmetric with respect to $\dphiX=0$. We show constraints on \cg and \ct obtained from $\tilde{A}_X$ in \Fig{fig:absasym_13tev}, where we obtain constrains which are very similar to the ones obtained using the full $\dphiX$ distributions. This means we can reliably construct observables that are unaffected by CP-even contributions but retain the best possible sensitivity to the CP-odd contributions.

A direct comparison between linear and quadratic contributions should be performed using the same analysis strategy for these two contributions. Since asymmetries are not suitable to study the quadratic contributions we use the full binned $\dphijj$ and $\dphill$ distributions in our analysis for linear and quadratic contributions. Including the quadratic dimension six contributions results in constraints shown in \Fig{fig:quad}\footnote{The stronger dependence of the $t\bar{t}h$ sample on \cg with respect to the linear case is a result of the \cg and \ct dependent branching ratio (see appendix) for $h\to b\bar{b}$. In the linear case the branching ratio retains the SM value since the linear CP-odd contributions vanish for this CP-even quantity.}. These constraints are much tighter than those obtained from the analysis in the linear case. \Fig{fig:linvsquadLHC} directly compares the limits obtained from the linear approximation and from the analysis which includes the quadratic contributions. 
This supports the previous point, highlighting that
the quadratic contributions are significant which results from the fact that they contribute to the total cross section in contrast to the linear contributions. This is also illustrated in \Fig{fig:linvsquadLHC} by comparing to bounds that are obtained from only the shape of the distributions discarding the information on the total cross section. The large effect of the quadratic contributions signals a violation of the perturbative constraint in \Eq{eq:val}. In other words, the stronger constraints in \Fig{fig:linvsquadLHC} rely on contributions that are perturbatively not under control and therefore should be treated with caution. In addition, including quadratic effects (which are CP-even) amounts to specific assumptions about the CP-even operators in the Higgs sector which cannot be disentangled straightforwardly anymore.

We explore how this situation changes as we moving to future colliders. Specifically, we study the two benchmark scenarios given in Tabs.~\ref{tab:scenario1} and \ref{tab:scenario2}. Scenario 1 can be considered as a worst-case scenario where the event selection efficiency $\epsilon_{t\bar{t}h}$ for $t\bar{t}h$ events and the systematic uncertainties do not improve and the integrated luminosity only moderately increases between the different colliders. Scenario 2 is an optimistic one where systematic uncertainties are reduced by a factor of about two when going to higher energies and $\epsilon_{t\bar{t}h}$ increases by a factor of two. Furthermore the integrated luminosity increases by an order of magnitude going from 13 TeV to 100 TeV in scenario 2. In order to obtain the efficiencies per bin for $hjj$ at 27 and 100 TeV we apply the same {\sc{Rivet}} analysis as for 13 TeV. The efficiencies are obtained as 14-20\%, 16-22\% and 19-25\% for 13, 27 and 100 TeV, respectively.  Since we already observe an analysis-inherent increase of efficiency for higher collider energies, we do not perform an additional rescaling in scenario 2 as it is done for the $t\bar{t}h$ efficiency. Hence, the $hjj$ efficiencies above are used for scenario 1 as well as scenario 2.
\begin{table}
\caption{Future collider scenario 1 used to study limits on \cg and \ct.}
\label{tab:scenario1}
{\renewcommand{\arraystretch}{1.6}
\renewcommand{\tabcolsep}{0.2cm}
\begin{tabular}{lcccc}
\hline
\hline
parameter & 13 TeV & 27 TeV & 100 TeV\\
\hline
$\mathcal{L}$ & 3/ab & 6/ab & 10/ab\\
\hline
\hline
parameter &\multicolumn{3}{c}{13,27,100 TeV}\\
\hline
$\dthproc{hjj}$ &             \multicolumn{3}{c}{10\%}\\
$\dsysflatproc{hjj}$ &        \multicolumn{3}{c}{10\%}\\
$\dsysshapeproc{hjj}$ &       \multicolumn{3}{c}{2.5\%}\\
$\epsilon_{t\bar{t}h}$ &      \multicolumn{3}{c}{2.5\%}\\
$\dthproc{t\bar{t}h}$ &       \multicolumn{3}{c}{10\%}\\
$\dsysflatproc{t\bar{t}h}$ &  \multicolumn{3}{c}{20\%}\\
$\dsysshapeproc{t\bar{t}h}$ & \multicolumn{3}{c}{1\%}\\
\hline
\hline
\end{tabular}}
\end{table}
\begin{table}
\caption{Future collider scenario 2 used to study limits on \cg and \ct.}
\label{tab:scenario2}
{\renewcommand{\arraystretch}{1.6}
\renewcommand{\tabcolsep}{0.2cm}
\begin{tabular}{lcccc}
\hline
\hline
parameter & 13 TeV & 27 TeV & 100 TeV\\
\hline
$\mathcal{L}$ & 3/ab & 10/ab & 30/ab\\
$\dthproc{hjj}$ & 10\% & 5\% & 2.5\%\\
$\dsysflatproc{hjj}$ & 10\% & 5\% & 2.5\%\\
$\dsysshapeproc{hjj}$ & 2.5\% & 1.5\% & 1.0\%\\
$\epsilon_{t\bar{t}h}$ & 2.5\% & 5\% & 10\%\\
$\dthproc{t\bar{t}h}$ & 10\% & 5\% & 2.5\%\\
$\dsysflatproc{t\bar{t}h}$ & 20\% & 10\% & 5\%\\
$\dsysshapeproc{t\bar{t}h}$ & 1\% & 1\% & 1\%\\
\hline
\hline
\end{tabular}}
\end{table}

The results of this study are shown in \Fig{fig:comp_energies} where the same analysis strategy for linear and quadratic contributions was applied. The increased centre-of-mass energy allows us to probe considerably higher energy scales, thus tightening the range of Wilson coefficients that can be considered to have a dominant effect from interference contributions (see \Eq{eq:valid}). However, the measurements become under increasing statistical control which will allow us to sharpen the exclusion. As can be seen in \Fig{fig:comp_energies}, the constraints from the linearised approach approximates quadratic exclusion. This shows that the quadratic contributions are considerably less relevant than we find for the LHC. This way the constraints at a 27 TeV \mbox{HE-LHC} will not only surpass the LHC, but will be more robust as well\footnote{We note that including additional channels and taking results from CMS into account even the constraints at a 13TeV High-Luminosity LHC could approach the perturbative limit. However, we do not quantify this statement here.}. As \Fig{fig:comp_energies} shows this is further strengthened at a 100 TeV machine, where the constraints for scenario 2 lie within the perturbative bounds given in \Eq{eq:valid}. Even in scenario 1 the bounds from linear terms are very close to those where quadratic terms are included.
Hence, we can probe Wilson coefficients in generic CP violating dimension six extensions in a perturbatively robust way, given our assumptions.

\begin{figure}[t!]
\includegraphics[height=6.5cm]{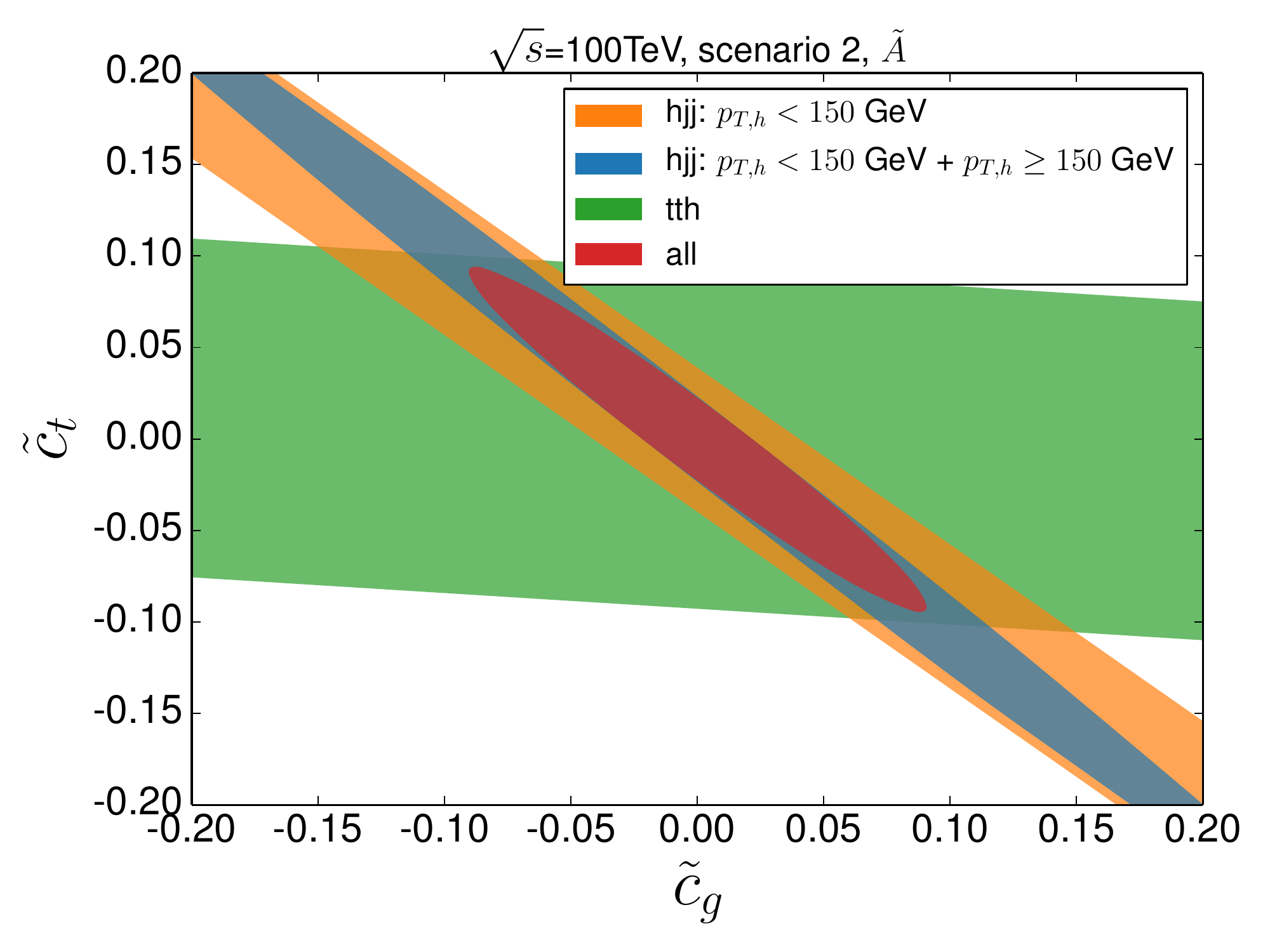}\hfill
\caption{\label{fig:absasym}Constraints for 95\% confidence level on \cg and \ct from different data sets at 100 TeV in scenario 2. Orange: only low-$p_{T,h}$ in \hjj, blue: combination of low $p_{T,h}$ and high $p_{T,h}$ events in \hjj, green: constraint from the \tth sample only, red: combination of all samples. The bounds were obtained using only the asymmetry observables defined in \Eq{eq:absasym}.
} 
\end{figure}

As a final remark we would like to point out that the bounds in \Fig{fig:comp_energies} are obtained using the full \dphill and \dphijj distributions. For completeness, we show constraints on \cg and \ct obtained from the asymmetry $\tilde{A}_X$ for a 100 TeV collider in scenario 2 in \Fig{fig:absasym}, which would be unaffected by CP-even contributions. As expected, we obtain constraints that are very similar to the ones obtained using the full $\dphiX$ distributions.

\section{Conclusions}
\label{sec:conc}

The interactions of the Higgs boson with the heaviest quarks in the SM
are motivated sources of CP violation. Analyses of top quark-related interactions that do not rely on particular Higgs final states and, consequently, are free of assumptions on
the Higgs decay couplings are largely limited to the dominant top-related Higgs production processes, \tth and \hjj.\footnote{In generic EFT deformations of the SM, there are different a priori sources of (C)P violation that we do not consider here.} Controlling competing effects from gluon-Higgs contact
interactions that might arise from additional heavy fermions are crucial in this context. The small statistics which is expected in the \tth channel with clean leptonic final states that enables a clean definition of sensitive observables based on the signed $\phi_{\ell\ell}$ limits the expected sensitivity as well as possibility to lift blind directions in the gluon-fusion related channels. 

Furthermore, and quite different from (C)P-even deformations of the SM, power-counting arguments for the effective interactions have a direct phenomenological consequence. While fully binned distributions provide a sensitive probe under all considerations of our work, for small CP-violating phases where we would expect SM interference-driven contributions to play a significant role in the limit setting, the decoupling of rate information seriously impacts the overall sensitivity of CP-analyses at the LHC. This is only partially mended at the high-energy LHC with 27 TeV as energy thresholds and expected statistics do not lead to a big enough improvement. While the precise specifications of a 100 TeV hadron collider are currently debated, the expected statistical improvement at such a machine locates the expected limit in a parameter region where the interference-driven interpretation starts saturating the EFT limit, i.e. power-counting assumptions do not impact the constraint quantitatively. The latter point is also supported by estimates of the EFT-related parameter validity ranges that are accessed through Monte Carlo simulations.

In summary we can state two main conclusions of our analysis. First,
CP violating effects in the top-Higgs sector can be extracted in a perturbatively robust way when measurements with high statistics are available. This can be realized by increased production cross sections at larger center-of-mass energies and increased integrated luminosities. We observe that for example at a 100 TeV collider \cg and \ct are constrained to a parameter region where quadratic dimension six contributions are considerably reduced resulting in perturbatively robust exclusion limits. This leaves the linear contribution as the dominant effect from dimension six operators. Second, since we consider CP-odd operators the linear contribution is indeed CP-odd while the quadratic contribution is CP-even which is difficult to disentangle from contributions of other CP-even operators. The fact that we can determine perturbatively robust results because only the linear contribution is dominant therefore also puts us in the position to cleanly study CP-odd SM deformations which otherwise would be intertwined with CP-even contributions.

\medskip
\noindent {\bf{Acknowledgments}} --- C.\ E.\ is supported by the IPPP Associateship scheme and by the UK Science and Technology Facilities Council (STFC) under grant ST/P000746/1.
P.\ G.\ is funded by the STFC under grant ST/P000746/1.
A.\ P.\ is supported by the Royal Society under grant UF160396 and by an IPPP Senior Experimental Fellowship. 
M.\ S.\ is funded by the STFC under grant  ST/P001246/1 and would like to thank the University of Tuebingen for hospitality and the Humboldt Society for support during the finalisation of parts of this work.

\appendix
\section{Branching ratios in the presence of squared dimension six contributions}
The operators \og and \ot add a pseudoscalar component to the following partial decay widths $\Gamma(h\to gg)$, $\Gamma(h\to\gaga)$ and $\Gamma(h\to \Zga)$ of the Higgs. Hence, the branching ratios \mbox{BR$(h\to\gaga)$} and \mbox{BR$(h\to b\bar{b})$} depend on \cg and \ct,
\begin{alignat*}{4}
\text{BR}(h\to b\bar{b}) & = & \frac{\GaSM^{b\bar{b}}}{\GaSM+\GaEFT}\,,\\[0.2cm]
\text{BR}(h\to\gaga) & = & \frac{\GaSM^{\gaga}+\ct^2\GaEFT^{\gaga}}{\GaSM+\GaEFT}
\end{alignat*}
with
\begin{multline*}
\GaEFT  =  \ct^2\left[\GaEFT^{\gaga}+\GaEFT^{\Zga}+\GaEFT^{gg,1}\right]\\+\cg^2\GaEFT^{gg,2}
 + \cg\ct\GaEFT^{gg,3}\,,
\end{multline*}
where $\GaSM$ is the total SM decay width of the Higgs, $\GaSM^{X}$ is the SM partial decay width into the final state $X$, $\GaEFT$ is the total decay width induced by the operators in \Eq{eq:ops} and $\GaEFT^{X}$ is the partial decay width into the final state $X$ due to dimension six operators. $\GaEFT^{\gaga}$, $\Gamma_{\text{dim.6.}}^{\Zga}$ and $\GaEFT^{gg,i}$ can be read off the pseudoscalar part of the decay widths given for example in \Ref{Djouadi:2005gj}:
\begin{equation*}
\begin{split}
\GaEFT^{\gaga} & =  \frac{G_F\alpha^2m_h^3}{72\sqrt{2}\pi^3}\left|A_{1/2}^A(\tau_h)\right|^2,\\
\GaEFT^{\Zga} & =  {\text{K}}_{\Zga}\frac{G_F^2m_W^2\alpha m_h^3(1-\frac{8}{3}s_w^2)^2}{16\pi^4c_w^2}\left(1-\frac{m_Z^2}{m_h^2}\right)^3\nonumber\\
&\times\left|B_{1/2}^A(\tau_h,\tau_Z)\right|^2,\\
\GaEFT^{gg,1} & =  {\text{K}}_{gg}\frac{G_F\alpha_s^2m_h^3}{64\sqrt{2}\pi^3}\left|A_{1/2}^A(\tau_h)\right|^2,\\
\GaEFT^{gg,2} & = {\text{K}}_{gg}\frac{G_F\alpha_s^2m_h^3}{16\sqrt{2}\pi^3}\,,\\
\GaEFT^{gg,3} & = {\text{K}}_{gg}\frac{G_F\alpha_s^2m_h^3}{16\sqrt{2}\pi^3}\text{Re}\left[A_{1/2}^A(\tau_h)\right]
\end{split}
\end{equation*}
with the loop functions
\begin{alignat*}{4}
A_{1/2}^A(\tau) & =  2f(\tau)/\tau\,,\\
B_{1/2}^A(\tau,\lambda) & =  \frac{1}{2(\tau-\lambda)}\left[f(\tau)-f(\lambda)\right]
\end{alignat*}
and
\begin{alignat*}{4}
f(\tau) & = & \left\{\begin{array}{ll}
\arcsin^2(\sqrt{\tau}) & \tau\le1\\
-\frac{1}{4}\left[\log\left(\frac{1+\sqrt{1-1/\tau}}{1-\sqrt{1-1/\tau}}\right)-i\pi\right]^2 & \tau > 1
\end{array}\right..\quad
\end{alignat*}
The functions' arguments are defined as
\begin{equation*}
\tau_h=\frac{m_h^2}{4m_t^2}\quad\text{and}\quad\tau_Z=\frac{m_Z^2}{4m_t^2}.
\end{equation*}
$c_w=\cos\theta_w$ and $s_w=\sin\theta_w$ where $\theta_w$ is the weak mixing angle. Finally, we rescale the partial
decay widths by the respective K factors \cite{Djouadi:2005gi}
\begin{equation*}
\begin{split}
{\text{K}}_{\Zga}&=1-\frac{\alpha_s}{\pi},\\
{\text{K}}_{gg}&=1+\frac{221}{12}\frac{\alpha_s(m_h)}{\pi}\nonumber\\
&+\left(\frac{\alpha_s(m_h)}{\pi}\right)^2\left(171.5-5\log\left(\frac{m_t^2}{m_h^2}\right)\right).
\end{split}
\end{equation*}
The K factor for a pseudoscalar decaying into $\gaga$ is one at NLO. The numerical value for the branching ratios as functions of \cg and \ct
used in the analysis are given by
\begin{equation*}
\begin{split}
\text{BR}(h\to b\bar{b}) & = \frac{0.577}{1+0.190\,\cg^2+0.397\,\cg\ct+0.208\,\ct^2}\,,
\end{split}
\end{equation*}
\begin{equation*}
\begin{split}
\text{BR}(h\to \gaga) & = \frac{0.00228+0.000413\,\ct^2}{1+0.190\,\cg^2+0.397\,\cg\ct+0.208\,\ct^2}
\end{split}
\end{equation*}
where the {\sc{Pdg}} \cite{Tanabashi:2018oca} values for $G_F$, $\alpha$, $m_Z$, the Higgs Cross Section Working Group \cite{Heinemeyer:2013tqa} values for the SM branching ratios of the Higgs
and $m_t=173$~GeV, $m_h=125$~GeV where used. We have cross-checked these results against an independent calculation 
using {\sc{FeynArts}}/{\sc{FormCalc}}/{\sc{LoopTools}}~\cite{Hahn:1998yk,Hahn:1999mt,Hahn:2000kx}.


\bibliography{paper.bbl}

\end{document}